\documentclass[aps,prd,10pt,notitlepage,nofootinbib,superscriptaddress,showkeys,showpacs]{revtex4-1}

\usepackage{amsmath,amssymb,amsthm,latexsym}
\usepackage[english]{babel}
\usepackage{color}
\usepackage{xspace}
\usepackage{graphicx}
\usepackage{pifont,dsfont}
\usepackage{marvosym}
\usepackage{slashed}
\usepackage[caption=false]{subfig}

\newcommand{\be}{\begin{equation}}
\newcommand{\ee}{\end{equation}}
\newcommand{\bqa}{\begin{eqnarray}}
\newcommand{\eqa}{\end{eqnarray}}
\newcommand{\bea}{\begin{eqnarray}}
\newcommand{\eea}{\end{eqnarray}}

\newcommand{\N}{\mathds{N}}

\newcommand{\R}{\mathds{R}}

\newcommand{\Z}{\mathds{Z}}

\newtheorem{lemma}{Lemma}

\newtheorem{definition}{Definition}
\newtheorem{theorem}{Theorem}
\newtheorem{remark}{Remark}
\newtheorem{corollary}{Corollary}
\newtheorem{proposition}{Proposition}

\DeclareMathOperator{\id}{id}

\DeclareMathOperator{\tr}{tr}

\DeclareMathOperator{\U}{U}
\DeclareMathOperator{\Pl}{Pl}
\DeclareMathOperator{\Mot}{Motzkin}
\DeclareMathOperator{\SIF}{SIF}

\begin{document}

\title{\Large \bf The calculation of expectation values in Gaussian random tensor theory via meanders}

\author{{\bf Valentin Bonzom}}\email{bonzom@lipn.univ-paris13.fr}
\affiliation{LIPN, UMR CNRS 7030, Institut Galil\'ee, Universit\'e Paris 13,
99, avenue Jean-Baptiste Cl\'ement, 93430 Villetaneuse, France, EU}
\author{{\bf Fr\'ed\'eric Combes}}\email{frederic.combes@ens-lyon.fr}
\affiliation{Perimeter Institute for Theoretical Physics, 31 Caroline St. N, ON N2L 2Y5, Waterloo, Canada}

\date{\small\today}

\begin{abstract}
\noindent A difficult problem in the theory of random tensors is to calculate the expectation values of polynomials in the tensor entries, even in the large $N$ limit and in a Gaussian distribution. Here we address this issue, focusing on a family of polynomials labeled by permutations, which naturally generalize the single-trace invariants of random matrix models. Through Wick's theorem, we show that the Feynman graph expansion of the expectation values of those polynomials enumerates meandric systems whose lower arch configuration is obtained from the upper arch configuration by a permutation on half of the arch feet. Our main theorem reduces the calculation of expectation values to those of polynomials labeled by stabilized-interval-free permutations (SIF) which are proved to enumerate irreducible meandric systems. This together with explicit calculations of expectation values associated to SIF permutations allows to exactly evaluate large $N$ expectation values beyond the so-called melonic polynomials for the first time.
\end{abstract}

\medskip


\maketitle

\section*{Introduction}

Random tensor theory \cite{TMReview} generalizes random matrix theory \cite{MMReview}. A tensor is a multi-dimensional array, here considered as a random variable. The observables are polynomials in the tensor entries invariant under some unitary transformations, and are the quantities whose expectation values we are interested in.

Tensor models have been first introduced in the context of quantum gravity \cite{Ambjorn3DTensors, SasakuraTensors, GrossTensors} (some tensor models, known as group field theories, provide a field theory framework for loop quantum gravity \cite{GFT}). Although the interest in tensor models has thus existed for a long time, it is only a few years ago that important progress was made, leading to the ability to solve some tensor models exactly in the limit of large tensor size $N$, \cite{Gur4, Uncoloring, CriticalColored}. This has had several applications: the discovery of new, non-local, perturbatively renormalizable field theories called tensorial field theories \cite{RenormalizableTGFT}, the analytical study of the continuum limit and critical phenomena of dynamical triangulations coupled to matter \cite{CriticalTensors} which confirmed the behavior of Euclidean Dynamical Triangulations observed numerically.

The main results concerning the large $N$ limit of random tensor models are naturally framed in probabilistic terms \cite{Universality, MelonsAreBP}. In particular, \cite{Universality} shows that the non-i.i.d. distributions considered in random tensor models become Gaussian at large $N$, which is a very strong universality result, while \cite{MelonsAreBP} proves that the contributions to the expectation value of an observable at large $N$ can be understood as random branched polymers, in particular in metric terms, this way finally confirming another expectation from the numerics.

While the theorems about random tensors are obviously probabilistic, the techniques hugely rely on combinatorics. The reason is that the expectation value of a polynomial $P(T)$ is expanded  using the Feynman expansion onto graphs,
\begin{equation*}
\langle P(T) \rangle = \sum_{\{\text{Feynman graphs}\}} \text{Feynman amplitudes},
\end{equation*}
($\langle \cdot \rangle$ denotes the expectation value), and the calculation therefore necessitates the understanding of the Feynman graphs and their associated amplitudes. The Feynman graphs of random tensor models are generically \emph{stranded} graphs (generalizing ribbon graphs) \cite{CombinatoricsRTMTanasa} and turn out to correspond to triangulations of pseudo-manifolds \cite{Ambjorn3DTensors} whose dimension is the number of indices of the tensor. This explains the longstanding difficulty of solving tensor models.

The breakthrough was to restrict to a particular class of models for which both the polynomials $P(T)$ and their Feynman graphs can be represented as \emph{regular edge-colored graphs} \cite{TMReview, Uncoloring} and the Feynman amplitudes depend on basic combinatorial properties of the graphs \cite{Gur4}. The set of colored graphs is much easier to handle than the set of stranded graphs, and the subset which dominate the large $N$ limit is in fact easily solvable \cite{Uncoloring, CriticalColored, LineColoredDAryTrees}. Actually, those regular colored graphs have recently been enumerated in the way that is relevant to tensor models in \cite{DoubleScalingSchaeffer}. Furthermore, it has been understood how to relax the colorability requirement in the case of three indices while still being able to solve the model \cite{LargeNMultiOrientable}. This is the first time a model based on stranded graphs has been solved (at large $N$).

In spite of the classification of \cite{DoubleScalingSchaeffer} (see also \cite{QuarticDoubleScaling} for a totally different approach dealing with a subset of colored graphs), there is no solution to the problem of evaluating the expectation value of an arbitrary polynomial, even at large $N$ only and even in a Gaussian distribution. The reason is that it is very difficult in general to find a characterization of the Feynman graphs contributing at large $N$ convenient enough so that they can be counted. The universality theorem of \cite{Universality} only asserts that the expectation value is (up to a constant) the same as in a Gaussian distribution, and provides an explicit calculation only for the so-called melonic polynomials whose Gaussian expectation value is just 1 (see also \cite{Uncoloring, SDE}). In this article, we are able for the first time to provide explicit calculation beyond the melonic case.

We focus on this issue: the exact calculation of expectation values of some polynomials of a \emph{Gaussian} random tensor \emph{at large $N$}.
\begin{itemize}
\item The polynomials we study are labeled by (one or two) permutations $\sigma\in\mathfrak{S}_n$, where $2n$ is the order of the polynomial. They are described in details in the Section \ref{sec:PolynomialsPermutations}. They generalize the single-trace invariants of random matrix models in the sense that the latter have a graphical representation with a single face while our polynomials have two faces superimposed in a non-trivial way.
\item We show in the Section \ref{sec:LargeNMeanders} that their Feynman expansion is an expansion onto meandric systems. A \emph{meandric system} \cite{Lacroix} consists of an upper and a lower planar arch configurations joined at the feet of the arches along a horizontal line so as to form closed non-intersecting curves crossing the horizontal line $2n$ times. The meandric systems contributing to the expectation value of a polynomial are such that the lower arch configuration is obtained from the upper one by applying the (one or two) permutations to (half of or all) the feet of the upper arches. These meandric systems are each counted with weight one, so that $\langle P_\sigma(T)\rangle$ simply enumerates them.
\item Our main theorems are in the Section \ref{sec:Factorization}. We prove that the expectation value of a polynomial can factorize as a product of expectation values of smaller bits. Those smaller bits are polynomials labeled by \emph{stabilized-interval-free} (SIF) permutations, which are permutations on $[1,n]$ which do not stabilize any subinterval $[i,j]$. Furthermore, the meandric systems contributing to their Feynman expansion are the \emph{irreducible} meandric systems, i.e. those which do not get disconnected after two cuts on the horizontal line.
\item The Section \ref{sec:Applications} offers applications of our factorization theorem to recover the numbers of meandric systems with $k$ components, for $k$ close to the order of the system (the number of crossings on the horizontal line). We also calculate the expectation values of polynomials of arbitrary degrees labeled by some SIF permutations.
\end{itemize}

As far as we know, this is the first time that the SIF permutations, studied in \cite{SIF}, are related to the irreducible meandric systems, which were introduced and studied in \cite{IrredMeanders}.

The meandric representation of the Feynman expansion connects the combinatorics of random tensor models to a well-known problem of enumerative combinatorics. Moreover, it turns out to be very convenient to study the expectation values and all our proofs are expressed using the meandric representation.

\emph{Notation.} Since only intervals of integers will be considered, we simply denote them with the standard notation $[a,b]$ of real intervals.

\section{Polynomials labeled by permutations} \label{sec:PolynomialsPermutations}

\subsection{Invariant polynomials in random tensor theory and their graphical representation}

Let $T$ be a rank $d$ tensor, with components $T_{a_1 \dotsb a_d}$, $a_i=1,\dotsc,N_i$ for $i=1,\dotsc,d$, and $\bar{T}$ its complex conjugate. Random tensor theory has been recently developed for $\U(N_1)\times \dotsb \times \U(N_d)$ invariant quantities, in the sense that the expectation values of invariant functions with respect to an invariant distribution on $T, \bar{T}$ are well-defined \cite{Uncoloring, Universality}. The algebra of invariant functions is generated by a set of polynomials labeled by connected edge-colored bipartite graphs of degree $d$. To build a polynomial $P_B(T,\bar{T})$ from a colored graph $B$, assign a $T$ to each white vertex, a $\bar{T}$ to each black vertex. For each edge with color $i\in\{1,\dotsc,d\}$, identify the indices $a_i$ in the position $i$ of the two tensors connected by the edge, and sum over $a_i$. This way all indices of all $T$ and $\bar{T}$ are contracted two by two in a $\U(N)$ invariant way. Some examples at $d=4$ are presented in the Figure \ref{fig:GraphsExamples}.

The expectation value of a polynomial $P_B$ is
\be \label{DefVEV}
\langle P_B \rangle = \int d\mu(T,\bar{T})\ P_B(T,\bar{T}),
\ee
where $d\mu$ is the joint distribution on the tensor entries. In the case $N_1=\dotsb=N_d=N$, the large $N$ limit of invariant distribution has been found in \cite{Universality} to be Gaussian, under some conditions that are typically satisfied in tensor models where $d\mu$ is a Gaussian measure perturbed with the exponential of invariant polynomials. In this case, the expectation values have a well-defined $1/N$ expansion, see a synthesis in \cite{Uncoloring}. The case where the sizes $N_i$ are different can lead to different behaviors in the large $N_i$ limits, detailed in \cite{new1/N} and \cite{LoopsColoredGraphs}.

We focus on the case $N_1=\dotsb=N_d=N$. The $1/N$ expansion of an expectation value reads
\be \label{VevExpansion}
\langle P_B \rangle = N^{\omega(B)} \sum_{k\in\N} N^{-k}\ C_k \underset{\text{large $N$}}{\simeq} N^{\omega(B)}\ [G^{(2)}]^{V/2}\ C_0^{(G)},
\ee
where the universality theorem for large random tensors \cite{Universality} allows to factorize the large $N$ dominant coefficient $C_0$ in terms of
\begin{itemize}
\item the large $N$, full covariance $G^{(2)}= \langle \sum_{\{a_i\}} T_{a_1\dotsb a_d} \bar{T}_{a_1\dotsb a_d}\rangle/N$ (the re-scaling makes $G^{(2)}$ of order $\mathcal{O}(1)$ at large $N$),
\item the half-number of vertices of $B$, $V/2$,
\item $C^{(G)}_0\in\N$ which is the leading order Gaussian average of $P_B$, counting the number of Wick pairings which are dominant at large $N$.
\end{itemize}
Evaluating expectation values therefore requires to calculate $G^{(2)}$, the observable scaling $\omega(B)$ and the amplitude $C^{(G)}_0$ for arbitrary colored graphs. This is obviously a difficult task. In this paper, we will focus on a Gaussian distribution with $G^{(2)}=1$, at $d=4$, and restrict attention to a specific family of invariant polynomials for which $i)$ $\omega(B)$ is easily found, $ii)$ more importantly $C^{(G)}_0$ counts the number of meanders such that the top and bottom arch configurations are related by a permutation.

\subsection{The family of interest}

To describe the family of observables we are interested in, it is useful to introduce the notion of faces.
\begin{definition}
{\rm (Graph faces).} Let $G$ be a bipartite connected edge-colored graph of degree $\Delta$ with colors in $\{1,\dotsc,\Delta\}$. A face of colors $(ij)$ is a connected closed subgraph with colors in $\{i,j\}$ only. In other words, we get the faces of colors $(ij)$ by erasing all edges with a different color and looking at the remaining connected pieces.
\end{definition}

In random matrix models, the polynomials associated to connected edge-colored bipartite graphs of degree 2 are the traces $\tr (MM^\dagger)^{V/2}$. The corresponding graphs are just closed cycles whose edge colors alternate 1 and 2 and with an even number of vertices $V$. Equivalently, they possess a single face with colors $(12)$.

\begin{figure}
\subfloat[The only 2-vertex, 4-colored graph.]{\begin{tabular}{c}\includegraphics[scale=.55]{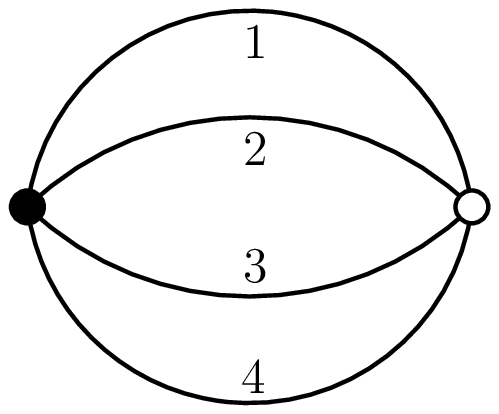}\end{tabular} \label{fig:2VertexGraph}}
\hspace{1.5cm}
\subfloat[The two 4-vertex graphs in $\mathcal{B}_2$.]{\begin{tabular}{c}\includegraphics[scale=.45]{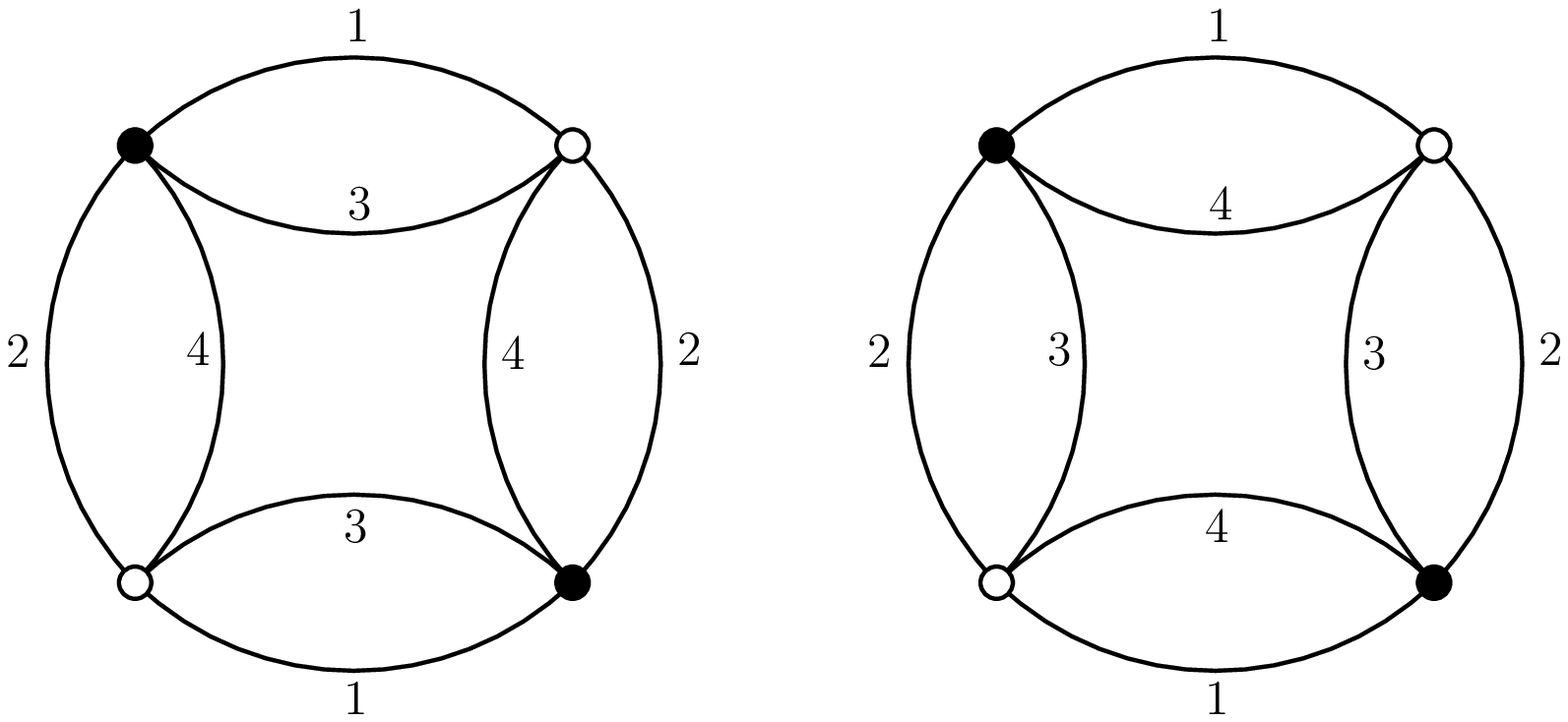}\end{tabular} \label{fig:4VertexGraph}}
\hspace{1.5cm}
\subfloat[A 10-vertex graph]{\begin{tabular}{c}\includegraphics[scale=.6]{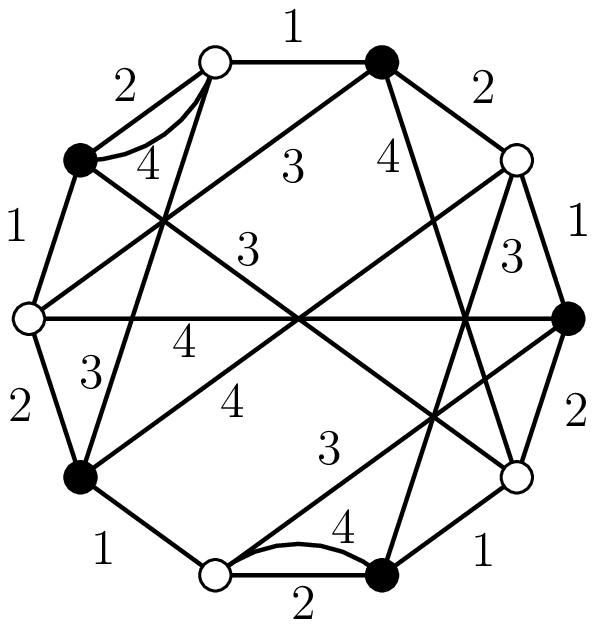}\end{tabular} \label{fig:10VertexGraph}}
\caption{Examples of graphs with a single face with colors $(12)$ and a single face with colors $(34)$. \label{fig:GraphsExamples}}
\end{figure}

In this article we consider connected bipartite 4-colored graphs (regular of degree 4) with a \emph{single} face of colors $(12)$ and a \emph{single} face of colors $(34)$, and we denote the set of such graphs on $2n$ vertices $\mathcal{B}_n$. We can represent any of them starting with the face of colors $(12)$ drawn as a $2n$-gon, and then glue to its vertices the face with colors $(34)$ (thereby typically creating crossings inside the $2n$-gon). Examples are provided in the Figure \ref{fig:GraphsExamples}.

Permutations are useful to label such graphs. The idea is to consider separately the face with colors $(12)$ and the face with colors $(34)$ with independently labeled vertices. It is then sufficient to say which white (respectively black) vertex of the face with colors $(34)$ is glued to which white (respectively black) vertex of the face with colors $(12)$. To do that more precisely, the following definition will be useful.
\begin{definition} \label{def:labeling}
{\rm (Face induced labeling).} Given a black (or white) vertex of reference labeled $1_\bullet$ (or $1_\circ$), a face with colors $(ab)$ induces a labeling $\{1_\bullet, 1_\circ, \dotsc,n_\bullet, n_\circ\}$ of the $2n$ vertices via the following rule: an edge of color $a$ connects the white vertex $j_\circ$ to the black vertex $j_\bullet$ and an edge of color $b$ connects the white vertex $j_\circ$ to the black vertex $(j+1)_\bullet$, for $j=1,\dotsc, n$ (with $n+1=1$).
\end{definition}

The cyclic group $\Z/n\Z$ acts on the labelings via the cyclic permutations $\Delta_p: i\in\{1,\dotsc,n\} \mapsto i+p \mod n \in\{1,\dotsc,n\}$, $p=0,\dotsc,n-1$, on both white and black vertices. Since there are $n$ possibilities for the vertex of reference, the action of $\Z/n\Z$ generates the whole set of labelings.

The following proposition characterizes graphs in $\mathcal{B}_n$ in terms of permutations.
\begin{proposition} \label{prop:BubbleLabeling}
A graph $B\in\mathcal{B}_n$ can be characterized by two permutations $\sigma_\circ, \sigma_\bullet \in\mathfrak{S}_n$, up to the left and the right actions of $\Z/n\Z$,
\be
\begin{aligned}
(\sigma_\circ, \sigma_\bullet) &\mapsto (\Delta_p\circ \sigma_\circ, \Delta_p\circ \sigma_\bullet), &\qquad &p=0,\dotsc,n-1,\\
(\sigma_\circ, \sigma_\bullet) &\mapsto (\sigma_\circ \circ \Delta_k, \sigma_\bullet\circ \Delta_k), &\qquad &k=0,\dotsc,n-1.
\end{aligned}
\ee
The graph is then denoted $B_{\sigma_\circ,\sigma_\bullet}$.
\end{proposition}

{\bf Proof.} We choose an arbitrary white vertex of reference, denoted $1_\circ$, and use it as the origin of a labeling of the other vertices induced by the face with colors $(12)$ (see the Definition \ref{def:labeling}). 
Then we choose a second white vertex of reference, denoted $1'_\circ$ and use it to get a second labeling of the vertices, this time induced by the face with colors $(34)$. 
This way, each white vertex gets two labels, say $i'_\circ$ from the face with colors $(34)$ and $j_{i\circ}$ from the face with colors $(12)$ (and $(i'_\bullet, j_{i\bullet})$ for black vertices, $i=1,\dotsc,n$). The permutations $\sigma_\circ, \sigma_\bullet$ are defined by
\be
\sigma_\circ(i'_\circ) = j_{i\circ},\qquad \text{and} \qquad \sigma_\bullet(i'_\bullet) = j_{i\bullet}.
\ee
They obviously depend on the choice of the vertices of reference $1_\circ, 1'_\circ$. If $p_\circ+1$ is chosen as the new vertex of reference $1_\circ$, the pair of permutations becomes $(\Delta_{-p} \circ \sigma_\circ, \Delta_{-p} \circ \sigma_\bullet)$. If the second vertex of reference $1'_\circ$ is chosen to be $(k+1)'_\circ$, the new permutations are $(\sigma_\circ \circ \Delta_k, \sigma_\bullet \circ \Delta_k)$.

The other way around, given two permutations $\sigma_\circ, \sigma_\bullet$ on $\{1,\dotsc,n\}$, we can reconstruct a graph. We draw the vertices and edges of colors 1,2 as a convex $2n$-gon and label the vertices as induced by the face with colors $(12)$ (from an arbitrary vertex of reference). Then we use $\sigma_\circ, \sigma_\bullet$ to add the colors 3 and 4. We connect the white vertex $\sigma_\circ(i)$ to $\sigma_\bullet(i)$ via an edge of color 3 and to $\sigma_\bullet(i+1)$ via an edge of color 4, for $i=1,\dotsc,n$. Obviously the same graph is obtained if $\Delta_{-p}\circ \sigma_\circ$ and $\Delta_{-p}\circ \sigma_\bullet$, or $\sigma_\circ \circ \Delta_k$ and $\sigma_\bullet \circ \Delta_k$, are used.
\qed

\begin{remark}
This labeling of the graphs by $(\sigma_\circ, \sigma_\bullet)$ uses as a reference the graph labeled by the identity on white and black vertices. It is a matrix-like observable, since two adjacent vertices are always connected by both an edge of color 1 and an edge of color 3, or by both an edge of color 2 and an edge of color 4,
\be
B_{\id,\id} = \begin{array}{c} \includegraphics[scale=.45]{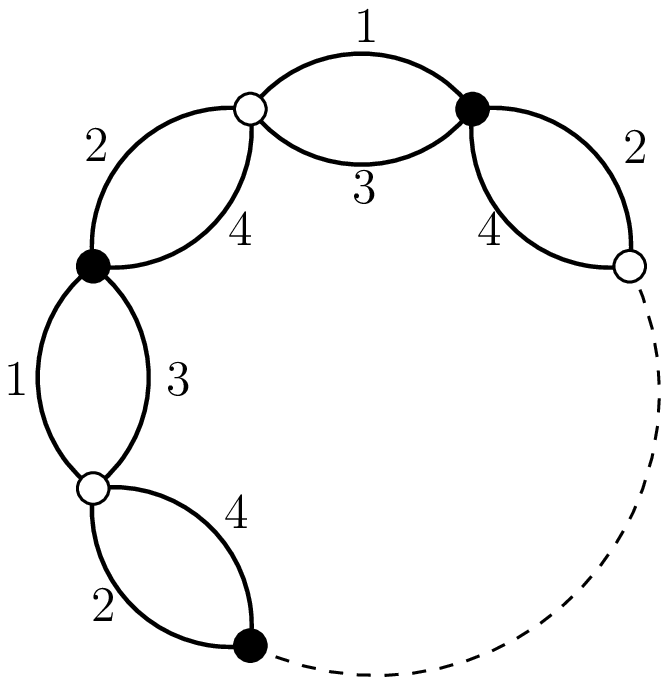} \end{array}
\ee
Therefore we could define fat-edges, labeled by the pair of colors $(13)$ or $(24)$ and corresponding to pairs of indices of $T$ and $\bar{T}$. In a Gaussian distribution, the corresponding polynomial of order $2n$ has the same expectation value at all orders as $\tr (MM^\dagger)^n$ for a random matrix $M$ of size $N^2\times N^2$ in a Gaussian distribution.

The effect of the permutations $\sigma_\circ, \sigma_\bullet$ is to move around the edges of colors 3 and 4 with respect to the matrix-like graph, by pulling out the vertices of the face with colors $(34)$ and dragging them to $\sigma_\circ(i), \sigma_\bullet(i)$.
\end{remark}

As an example, the 10 vertex graph in Figure \ref{fig:10VertexGraph} can be labeled as follows, (we have used the color code 1=red, 2= black, 3=green, 4=blue),
\be
B_{\sigma_\circ,\sigma_\bullet} = \begin{array}{c} \includegraphics[scale=.5]{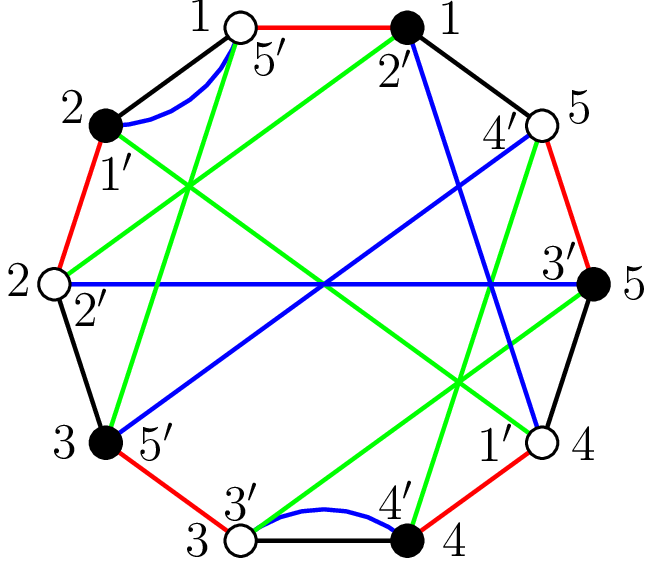} \end{array},\quad \text{with } \left\{ \begin{aligned} \sigma_\circ &= (145)(2)(3),\\
\sigma_\bullet &= (12)(35)(4). \end{aligned} \right.
\ee

\begin{proposition}
Let us denote $\mathcal{B}_{n,\circ}$ the set of graphs as in $\mathcal{B}_n$ but equipped in addition with a marked white vertex. Then there is a bijection between $\mathcal{B}_{n,\circ}$ and $\mathfrak{S}_n \times \mathfrak{S}_{n-1}$.
\end{proposition}

{\bf Proof.} It parallels the proof of the Proposition \ref{prop:BubbleLabeling}, using the marked vertex as $1_\circ = 1'_\circ$ to set the labels unambiguously. Thus we get $\sigma_\bullet$ and $\sigma_\circ$ as before, but since $1_\circ = 1'_\circ$, we always get $\sigma_\circ(1)=1_\circ$ and then $\sigma_\circ \in\mathfrak{S}_{n-1}$.
\qed

The interest of this Proposition lies in the fact that the Schwinger-Dyson equations, a set of algebraic equations which relate the expectation values of all polynomials to one another, are labeled by regular edge-colored graphs with a marked vertex, and are therefore well-labeled by $\mathfrak{S}_n \times \mathfrak{S}_{n-1}$. Instead of Schwinger-Dyson equations, we will use the Feynman expansion, but more comments on them can be found in the Conclusion.

\subsection{$1/N$ expansion of Gaussian expectation values}

The Gaussian measure we consider is
\be
d\mu_G(T,\bar{T}) = \frac1Z\ e^{-N^2\,T\cdot \bar{T}}\ dT\,d\bar{T},
\ee
where $T\cdot \bar{T} = \sum_{a_1, a_2, a_3, a_4} T_{a_1 a_2 a_3 a_4} \bar{T}_{a_1 a_2 a_3 a_4}$, and $Z$ is the normalization.

Note that the power of $N$ in the Gaussian is not the one usually considered for rank 4 tensors (that would be $N^3$ instead of $N^2$). However, it has been shown in \cite{new1/N} that $N^2$ makes sense too (and for non-Gaussian joint distributions, the large $N$ limit is out of the range of applicability of the universality theorem, so non-Gaussian large $N$ limits can be observed). Anyways, when the joint distribution is Gaussian as in our case, the scaling is not really relevant. Indeed, if one introduces $S = T/\sqrt{N}$, the Gaussian becomes $e^{-N^3 S\cdot \bar{S}}$, and the expectation value of $P_B(S, \bar{S})$ simply differs from that of $P_B(T, \bar{T})$ by a factor $N^{V/2}$, $V$ being the degree of $P_B$ (number of vertices of the corresponding colored graph). Since the family $\mathcal{B}_n$ of observables we are going to study has a uniform scaling, i.e. independent of the number of vertices \footnote{However, the leading order coefficient $C_0$ might vanish.}, with $N^2$ in the Gaussian, this choice appears as the most natural one.

Consider a graph $B_{\sigma_\circ, \sigma_\bullet}\in\mathcal{B}_{n}$ labeled by the permutations $\sigma_\circ, \sigma_\bullet$ and the corresponding polynomial $P_{\sigma_\circ, \sigma_\bullet}$ of degree $n$ in $T$ and in $\bar{T}$. According to Wick's theorem, the Gaussian average of $P_{\sigma_\circ, \sigma_\bullet}$ has an expansion onto Wick pairings,
\be
\langle P_{\sigma_\circ, \sigma_\bullet}(T,\bar{T}) \rangle = \sum_{\text{Wick pairings $\pi$}} N^{\Omega(\sigma_\circ,\sigma_\bullet,\pi)}.
\ee
A \emph{Wick pairing} is a way of associating to each $T$ a different $\bar{T}$. Using a labeling induced by the face with colors $(12)$, it can therefore be seen as a permutation $\pi\in\mathfrak{S}_n$ which associates to each label in $\{1_\circ,\dotsc,n_\circ\}$ a label in $\{1_\bullet, \dotsc, n_\bullet\}$. It can be represented graphically via additional edges, say carrying the fictitious color 0, between the vertices labeled with $i_\circ$ and $\pi(i)_\bullet$.

The labeling induced by the face with colors $(12)$ together with $\sigma_\circ$ induces a second labeling, $\{1'_\circ,1'_\bullet,\dotsc,n'_\circ,n'_\bullet\}$, compatible with the face of colors $(34)$. The label $i'_\circ$ is given to the vertex with label $\sigma_\circ(i)$ and the label $i'_\bullet$ goes to the vertex with label $\sigma_\bullet(i)$. This is the labeling induced by the face with colors $(34)$ with the vertex labeled $\sigma_\circ(1)$ chosen as the reference $1'_\circ$. The Wick pairing is also a permutation on this second set of labels: it connects $i'_\circ\in\{1'_\circ,\dotsc, n'_\circ\}$ to $[\sigma_\bullet^{-1}\circ \pi \circ \sigma_\circ (i)]'_\bullet \in \{1'_\bullet,\dotsc,n'_\bullet\}$.

The graph $B_{\sigma_\circ,\sigma_\bullet}$ dressed with the additional lines of color 0 representing the Wick pairing $\pi$ (they are called \emph{propagators} in quantum field theory) is denoted $G_{\sigma_\circ, \sigma_\bullet,\pi}$. It is a connected, bipartite, edge-colored graph with five colors. For instance, there are two such graphs in the expansion of the expectation value of the graph of the Figure \ref{fig:4VertexGraph},
\be
\left\langle \begin{array}{c} \includegraphics[scale=.4]{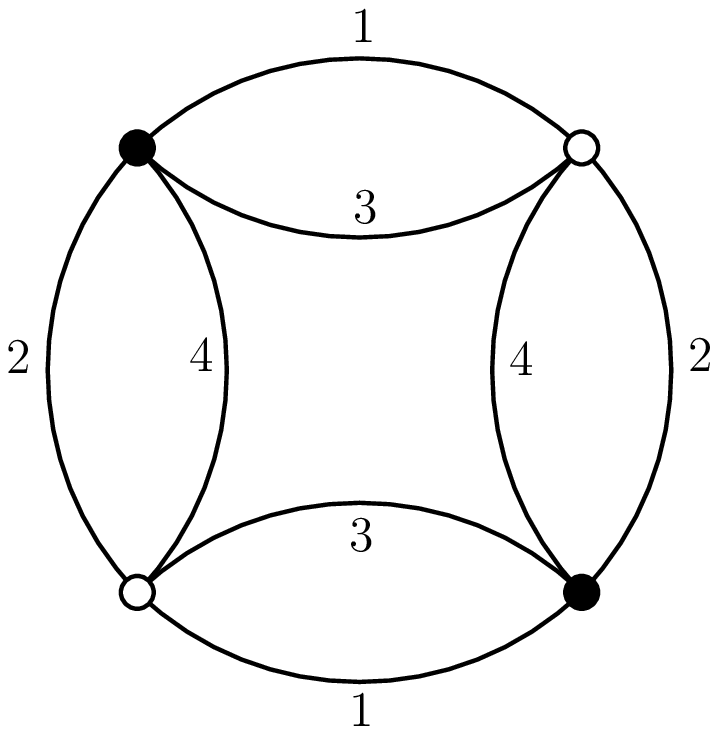} \end{array} \right\rangle = \begin{array}{c} \includegraphics[scale=.4]{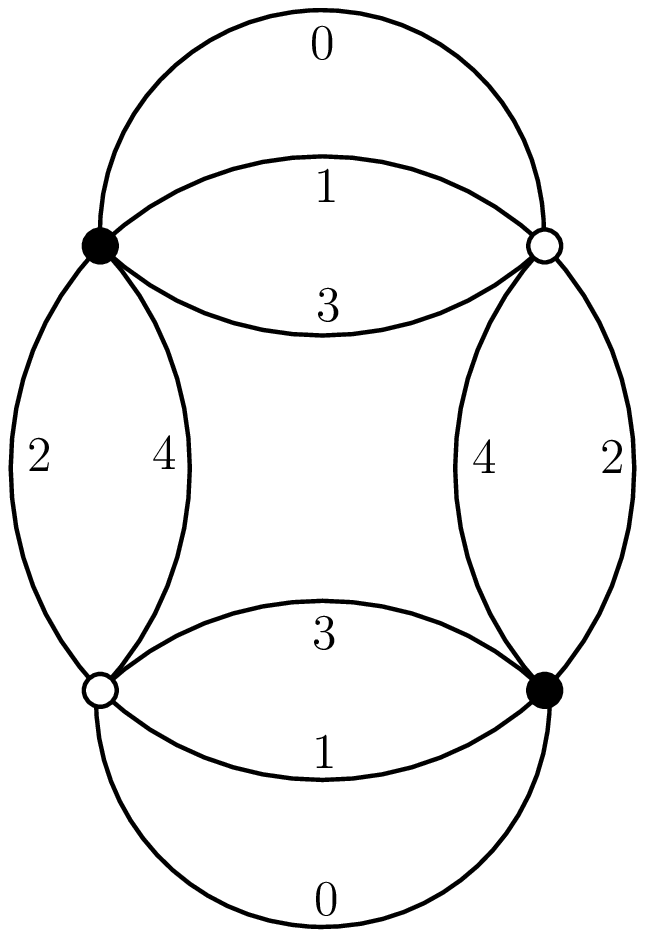} \end{array} + \begin{array}{c} \includegraphics[scale=.4]{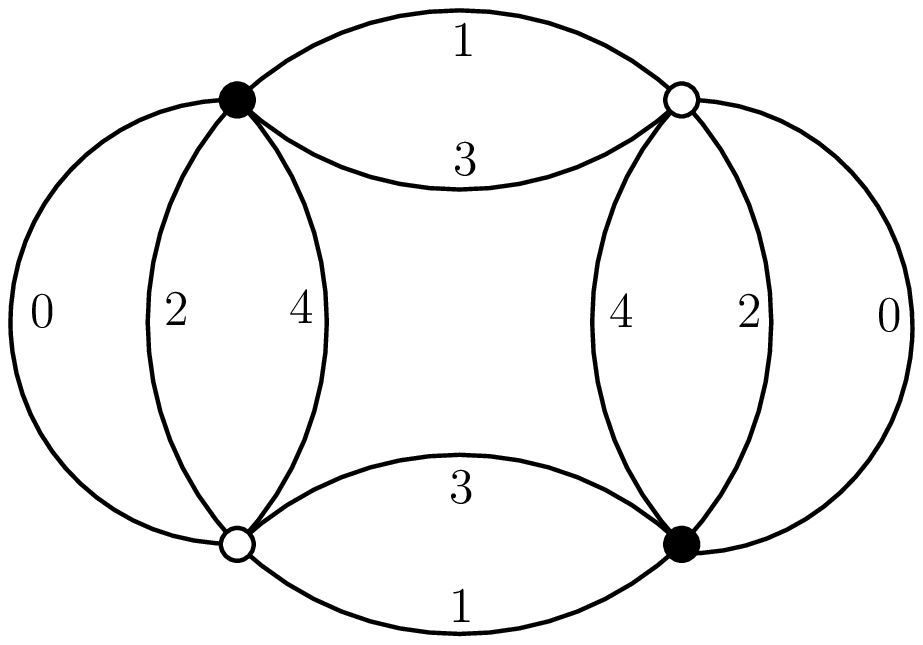} \end{array}
\ee

In addition to faces of colors $(ij)$, $i,j=1,2,3,4$, it also has faces with colors $(0i)$, $i=1,2,3,4$. It turns out that the exponent of $N$ associated to a Wick pairing $\pi$ can be expressed as \cite{Uncoloring, new1/N}
\be
\Omega(\sigma_\circ, \sigma_\bullet,\pi) = \sum_{i=1}^4 f_{0i} - 2 \ell_0,
\ee
where $f_{0i}$ is the number of faces with colors $(0i)$ and $\ell_0$ is the number of edges with color 0, $\ell_0 = n$. In the following $\ell_i$ is the number of edges of color $i$, with $\ell_i = n$ obviously. We consider the subgraph obtained by erasing from $G_{\sigma_\circ, \sigma_\bullet,\pi}$ the colors 3 and 4. It is a connected bipartite 3-colored graph with vertices of degree 3 and therefore represents the cell decomposition of a topological orientable surface whose genus $g_{12}(\sigma_\circ, \sigma_\bullet,\pi)$ is given by the classical formula
\be
2-2g_{12}(\sigma_\circ, \sigma_\bullet,\pi) = \underbrace{f_{12}+f_{01}+f_{02}}_{\text{total $\#$ of faces}} - (\underbrace{\ell_1+\ell_2+\ell_0}_{\text{total $\#$ of edges}}) + 2n = 1+f_{01}+f_{02} - \ell_0.
\ee
In the second equality, we have used $f_{12}=1$ and $\ell_0 = \ell_1=\ell_2=n$. Everything works similarly for the subgraph with the colors 1 and 2 erased. It is a graph with colors 0,3,4, dual to a triangulation of a topological surface whose genus is given by
\be
2-2g_{34} = 1+f_{03}+f_{04} - \ell_0.
\ee

Therefore the exponent of $N$ coming from a Wick pairing $\pi$ is
\be
\Omega(\sigma_\circ, \sigma_\bullet,\pi) = 2 - 2g_{12}(\sigma_\circ, \sigma_\bullet,\pi) - 2g_{34}(\sigma_\circ, \sigma_\bullet,\pi),
\ee
and we get a doubled 2D topological expansion,
\be \label{VevTopologicalExpansion}
\langle P_{\sigma_\circ, \sigma_\bullet}(T, \bar{T}) \rangle = N^2 \sum_{\text{Wick pairings $\pi$}} N^{-2g_{12}(\sigma_\circ, \sigma_\bullet, \pi)-2g_{34}(\sigma_\circ, \sigma_\bullet, \pi)}.
\ee
In particular, the coefficient $C_k$ of the $1/N$ expansion in \eqref{VevExpansion} is the number of Wick pairings such that the sum of genera $g_{12}(\sigma_\circ, \sigma_\bullet,\pi)+g_{34}(\sigma_\circ, \sigma_\bullet,\pi)$ is $k/2$.

Notice that it is possible to express the topological quantities $g_{12}(\sigma_\circ, \sigma_\bullet, \pi), g_{34}(\sigma_\circ, \sigma_\bullet, \pi)$ in terms of properties of the permutations $\pi$ and $\sigma_\bullet^{-1}\circ \pi\circ \sigma_\circ$ (and the number of vertices). Indeed, let us start at a vertex $i_\circ$ and list the vertices we meet when following the edges of colors 0 and 1:
\be
i_\circ \leftrightarrow_1 i_\bullet \leftrightarrow_0 \pi^{-1}(i)_\circ \leftrightarrow_1 \pi^{-1}(i)_\bullet \leftrightarrow_0 (\pi^{-1}\circ\pi^{-1})(i)_\circ \leftrightarrow_1 (\pi^{-1}\circ\pi^{-1})(i)_\bullet\dotsb,
\ee
where $\leftrightarrow_a$ means there is an edge with color $a$. Therefore the number of faces with colors $(01)$ is
\be
f_{01} = z(\pi^{-1}),
\ee
where $z(\pi)$ denotes the number of cycles of the permutation. Similarly, following the colors 0 and 2, one meets the following vertices $(i_\bullet \leftrightarrow_2 (i+1)_\circ \leftrightarrow_0 \pi(i+1)_\bullet \leftrightarrow_2 (\pi(i+1)+1)_\bullet \leftrightarrow_0 \pi(\pi(i+1)+1)_\circ \dotsb$, so that
\be
f_{02} = z(\Delta_1\circ \pi).
\ee
(Remember that $\Delta_1$ is the cyclic permutation $i\mapsto i+1 \mod n$.) With the same reasoning,
\be
f_{03} = z(\sigma_\circ^{-1} \circ \pi^{-1} \circ \sigma_\bullet),\qquad f_{04} = z(\Delta_1 \circ \sigma_\bullet^{-1} \circ \pi \circ \sigma_\circ).
\ee

\section{The meandric representation of large $N$ Gaussian expectation values} \label{sec:LargeNMeanders}

\subsection{Gaussian expectation values as the enumeration of meandric systems}

\subsubsection{Meandric systems}

We first give the classic informal picture of a meander. Consider a river, oriented from west to east, with $2n$ bridges. A meander is a closed, self-avoiding road which crosses all the bridges. A meandric system with $k$ roads is a set of $k$ non-intersecting meanders. A more formal definition is the following one.

\begin{definition}
{\rm (Meanders and meandric systems).} A meander of order $n$ is a closed, self-avoiding curve (the road) which crosses an infinite oriented horizontal line (the river) exactly $2n$ times (the bridges). The number of meanders of order $n$ is denoted $M_n$.

A meandric system of order $n$ with $k$ components is a set of $k$ non-crossing meanders all intersecting the same horizontal line, exactly $2n$ times in total. The number of meandric systems with $k$ components and of order $n$ is denoted $M_n^{(k)}$. Two systems are equivalent if there is a homeomorphism of the plane mapping one to the other.
\end{definition}

Another representation of the problem of counting the number of meanders is the problem of calculating the entropy associated to compact foldings of a polymer on the plane \cite{ArchStat, Meanders:ExactAsymptotics}.

\begin{figure}
\includegraphics[scale=.8]{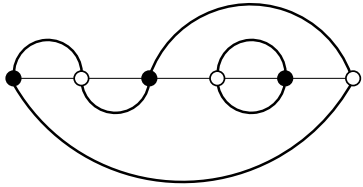}
\caption{\label{fig:MeandricSysExample} A meandric system of order 3 with two connected components.}
\end{figure}

We will use the canonical representation, where the river is oriented from west to east, has $2n$ marked vertices (black and white ones), the segments of the roads above and under the river are represented as semi-circular arches (caps and cups) whose feet are the vertices. An example is provided in the Figure \ref{fig:MeandricSysExample}.

\subsubsection{Graphical re-encoding of Wick pairings}

The contributions which dominate the large $N$ limit in the Equation \eqref{VevTopologicalExpansion} are the graphs $G_{\sigma_\circ,\sigma_\bullet, \pi}$ such that the subgraphs with colors 0,1,2 and with colors 0,3,4 both have vanishing genus. As usual, this can be formulated as a planarity criterion. First, draw the face with colors $(12)$ as a convex $2n$-gon and the lines of color 0 on the exterior region, joining the vertices with labels $i_\circ$ and $\pi(i)_\bullet$. The genus $g_{12}(\sigma_\circ,\sigma_\bullet,\pi)$ is zero if and only if this graph is planar. Then proceed similarly with the face with colors $(34)$, which has to be "unfolded". Draw it as a convex $2n$-gon with vertex labels $\{1'_\circ,1'_\bullet,\dotsc,n'_\circ,n'_\bullet\}$. Then the lines of color 0 connect $i'_\circ$ to $[\sigma_\bullet^{-1} \circ \pi\circ \sigma_\circ (i)]'_\bullet$. Draw them on the exterior of the $2n$-gon. It comes that the genus $g_{34}(\sigma_\circ, \sigma_\bullet, \pi)$ vanishes if and only if this graph is planar.

The drawback of this representation is that we actually need two separate drawings in order to draw both the faces with colors $(12)$ and with colors $(34)$ as $2n$-gons. To improve the situation, we can draw the lines of color 0 inside the $2n$-gon with colors $(34)$, which does not change the equivalence between planarity and vanishing genus. Then we can identify the two $2n$-gons, to get a single one, with edges of color 0 on the exterior, representing the permutation $\pi$, and a copy of the Wick pairing on the inside representing the permutation $\sigma_\bullet^{-1}\circ \pi\circ \sigma_\circ$. Finally we can cut the $2n$-gon and stretch it horizontally. This way we obtain a horizontal line with semi-circular arches on the upper half plane and on the lower half-plane.

In the following it will be useful to identify permutations with arch configurations.
\begin{definition} \label{def:PermArch}
{\rm (Permutations and arch configurations).} A permutation $\rho\in\mathfrak{S}_n$ can be represented as a (most of the time non-planar) arch configuration on the set of $2n$ ordered vertices $(1_\bullet, 1_\circ,\dotsc,n_\bullet, n_\circ)$, by ordering the vertices on a horizontal line, from left to right, and drawing arches between $i_\circ$ and $\rho(i)_\bullet$, $i=1,\dotsc,n$. The arches can be drawn all in the upper or lower half-plane. The other way around, any arch configuration gives rise to a permutation.

A permutation $\rho$ is said to be planar, $\rho\in\Pl\mathfrak{S}_n$, if its arch configuration is planar.
\end{definition}

Step by step, our new representation of a Wick pairing $\pi\in\mathfrak{S}_n$ on a graph $B_{\sigma_\circ, \sigma_\bullet}\in\mathcal{B}_n$ is obtained in the following way.
\begin{itemize}
\item Draw a horizontal line with alternating black and white vertices which get the labels $(1_\bullet,1_\circ,\dotsc,n_\bullet,n_\circ)$ from west to east.
\begin{equation*}
\begin{array}{c}
\includegraphics[scale=.85]{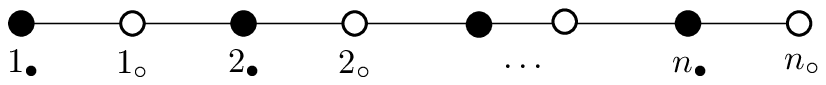}
\end{array}
\end{equation*}
\item The permutation $\pi$ is represented in the upper half-plane by semi-circular arches connecting $i_\circ$ to $\pi(i)_\bullet$, $i=1,\dotsc,n$. The genus $g_{12}(\sigma_\circ, \sigma_\bullet, \pi)$ vanishes if and only if the arch configuration is planar (note that this is however independent of $\sigma_\circ$ and $\sigma_\bullet$).
\begin{equation*}
\begin{array}{c}
\includegraphics[scale=.85]{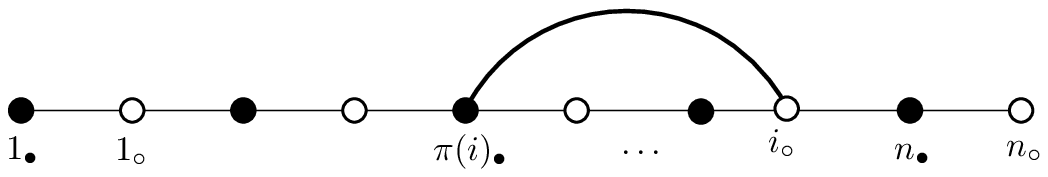}
\end{array}
\end{equation*}
\item The permutation $\sigma_\bullet^{-1}\circ \pi\circ \sigma_\circ$ is represented in the lower half-plane by semi-circular arches connecting $i_\circ$ to $[\sigma_\bullet^{-1}\circ \pi\circ \sigma_\circ(i)]_\bullet$, $i=1,\dotsc,n$, and planarity of the arch configuration is equivalent to $g_{34}(\sigma_\circ, \sigma_\bullet, \pi)=0$.
\begin{equation*}
\begin{array}{c}
\includegraphics[scale=.85]{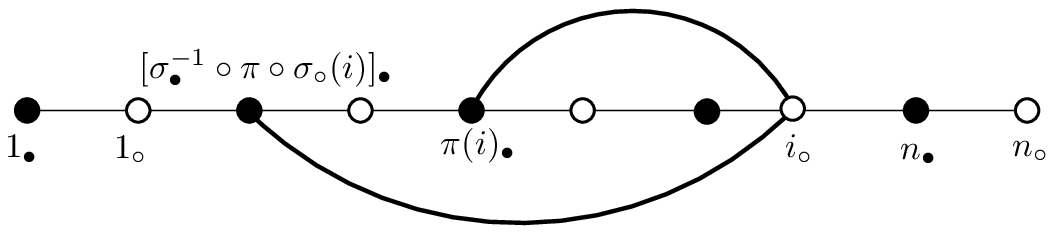}
\end{array}
\end{equation*}
\item We end up with two arch configurations, which form closed roads winding across a river. In the large $N$ limit, only the planar arch configurations survive, which are exactly meandric systems of order $n$.
\end{itemize}

\begin{proposition}
Let $\mathcal{M}_{\sigma_\circ, \sigma_\bullet}$ be the set of meandric systems such that if $\pi\in\Pl\mathfrak{S}_n$ is the upper arch configuration, then $\sigma_\bullet^{-1}\circ \pi\circ \sigma_\circ$ is planar too and represents the lower arch configuration. We have thus shown that
\be \label{VEVMeandricExpansion}
\langle P_{\sigma_\circ, \sigma_\bullet} \rangle \underset{\text{large $N$}}{=} \vert \mathcal{M}_{\sigma_\circ, \sigma_\bullet} \vert.
\ee
\end{proposition}

In matrix model, graphs corresponding to the observables $\tr (MM^\dagger)^n$ have a single face (with colors $(12)$) and by following the same reasoning as above, a Wick pairing is a permutation $\pi\in\mathfrak{S}_v$ or equivalently an arch configuration. The number of planar permutations, describing a planar arch configuration, is the Catalan number $C_n$ and indeed the large $N$ evaluation
\be
\langle \tr (MM^\dagger)^n\rangle = C_n,
\ee
is very well-known. However, in our case, the presence of a second face, with colors $(34)$, supported on the same set of vertices, makes the evaluation much more involved and explains the need for two arch configurations, and hence meandric systems in the large $N$ limit. Given $\sigma_\circ, \sigma_\bullet$ and a planar upper arch configuration encoded by $\pi$, the lower arch configuration, representing $\sigma_\bullet^{-1} \circ \pi \circ \sigma_\circ$, is typically not planar\ldots

This provides a very simple bound: the number of Wick pairings which contribute at large $N$ is bounded by the number of planar arch configurations,
\be \label{CatalanBound}
\langle P_{\sigma_\circ, \sigma_\bullet} \rangle \leq C_n.
\ee

\subsection{Meandric permutations}

Meandric systems can be described by permutations called \emph{meandric permutations} \cite{Lacroix}. First consider a labeling of the vertices of the horizontal line, say from left to right, $1,2,\dotsc, 2n$, the odd vertices being the black ones and the even vertices being the white ones. The roads are also oriented such that they go from bottom to top at each black vertex. The meandric permutation $\rho$ is defined as a product of disjoint cycles, one for each road whose cycle is obtained by listing the vertex labels encountered along the road.

While there is a one-to-one correspondence between meandric systems and meandric permutations, it is very hard to identify the meandric permutations. A necessary condition is the following: if $\rho$ has $k$ cycles, then $\rho^2$ has $2k$ cycles, $k$ of them on the odd labels $\{1,3,\dotsc,2n-1\}$ and the others on the even labels $\{2,4,\dotsc,2n\}$.

It is easy to relate $\rho$ to our permutations $(\sigma_\circ, \sigma_\bullet, \pi)$.
\begin{proposition}
Consider a meandric system in $\mathcal{M}_{\sigma_\circ, \sigma_\bullet}$ with $\pi\in\Pl\mathfrak{S}_n$ as the upper arch configuration. Let $\rho\in \mathfrak{S}_{2n}$ denote the corresponding meandric permutation. Then
\be
\begin{aligned}
\rho^2(2i-1) &= \left(\sigma_\bullet^{-1} \circ \pi \circ \sigma_\circ \circ \pi^{-1}\right) (i),&& \\
\rho^2(2i) &= \left(\pi^{-1} \circ\sigma_\bullet^{-1} \circ \pi \circ \sigma_\circ \right) (i), &&
\text{for $i=1,\dotsc,n$.}
\end{aligned}
\ee
\end{proposition}
The proof simply tracks the labels along the oriented roads. It has the following interesting consequence.
\begin{corollary} \label{cor:NumberCycles}
A meandric system in $\mathcal{M}_{\sigma_\circ, \sigma_\bullet}$ with $\pi\in\Pl\mathfrak{S}_n$ as the upper arch configuration has exactly $z(\sigma_\bullet^{-1} \circ \pi \circ \sigma_\circ \circ \pi^{-1})$ closed curves, where $z$ denotes the number of cycles. In the case $\sigma_\circ=\id$ (respectively $\sigma_\bullet=\id$), this reduces to $z(\sigma_\bullet)$ (respectively $z(\sigma_\circ)$) and is therefore independent of the Wick pairing $\pi$.
\end{corollary}

\subsection{From meanders to Gaussian expectation values}

We have shown that the expectation values of polynomials in a specific family can be evaluated as a number of meandric systems. We would also like to know whether while exploring the whole family of polynomials we encounter all meandric systems. It turns out to be the case, in the sense that from a meandric system and a given $\sigma_\circ$, it is possible to reconstruct a graph $B_{\sigma_\circ, \sigma_\bullet}$ and a Wick pairing $\pi$.
\begin{proposition}
For any fixed $\sigma_\circ\in\mathfrak{S}_n$, there is a one-to-one correspondence between meandric systems of order $n$ and elements of $\{\mathcal{M}_{\sigma_\circ, \sigma_\bullet}\}_{\sigma_\bullet\in\mathfrak{S}_n}$.
\end{proposition}

{\bf Proof.} An element of $\{\mathcal{M}_{\sigma_\circ, \sigma_\bullet}\}_{\sigma_\circ, \sigma_\bullet}$ is a meandric system characterized by $\sigma_\circ$ (fixed), $\sigma_\bullet$ and an arch configuration $\pi$. We thus have to show that there exist a unique $\pi$ and a unique $\sigma_\bullet$ for any meandric system. The vertices along the horizontal line are labeled from left to right, $(1_\bullet, 1_\circ, \dotsc, n_\bullet, n_\circ)$. By following the top arches we simply read $\pi$: $\pi(i)$ is the label of the black vertex connected to $i_\circ$ by an upper arch (see the Definition \ref{def:PermArch}). We proceed similarly with the arches in the lower half-plane: $\sigma_\bullet^{-1}\circ \pi\circ \sigma_\circ (i)$ is the label of the black vertex connected to $i_\circ$ by a lower arch. Since $\sigma_\circ$ and $\pi$ are known, this defines $\sigma_\bullet$.
\qed

\begin{corollary}
This immediately implies
\begin{align}
&\forall \sigma_\circ\in\mathfrak{S}_n &&\sum_{\sigma_\bullet\in\mathfrak{S}_n} \langle P_{\sigma_\circ, \sigma_\bullet} \rangle = \sum_{k\geq1} M_n^{(k)}= C_n^2,\\
&\text{and} &&\sum_{\substack{\sigma_\bullet\in\mathfrak{S}_n \\ z(\sigma_\bullet) = k}} \langle P_{\id, \sigma_\bullet} \rangle = M^{(k)}_n, \label{SumFixedNumberCycles}
\end{align}
(since the total number of meandric systems of order $n$ is the square of the Catalan number $C_n$).
\end{corollary}

In the Equation \eqref{SumFixedNumberCycles}, we have specialized $\sigma_\circ=\id$ so that the number of closed loops of the meandric systems are precisely the number of cycles of $\sigma_\bullet$ (Corollary \ref{cor:NumberCycles}).

\section{Factorization on stabilized-interval-free permutations} \label{sec:Factorization}

In this section we restrict attention to the case $\sigma_\circ=\id$, and to simplify the notation we only write $\sigma_\bullet = \sigma$ explicitly, like $B_{\sigma}, P_{\sigma}$ and so on.

In the trivial case we have
\be
\langle P_{\id}(T, \bar{T}) \rangle = M_n^{(n)} = C_n.
\ee
From the point of view of Wick's theorem, the $C_n$ contributions come from the fact that the two 3-colored graphs formed by the all vertices and the edges of colors 0,1,2, and those of colors 0,3,4 are the same. Therefore the expectation value is the same as for a single-trace invariant in a Gaussian matrix model. In terms of meandric systems, this means that all the $C_n$ planar arch configurations on the upper half-plane are trivially reflected in the lower half-plane with respect to the horizontal line, by the trivial permutation on the vertices: the top and bottom arch configurations are the same. They correspond to all the meandric systems with exactly $n$ loops on $2n$ vertices.

In the general case, we have the bound \eqref{CatalanBound}, but we would like to evaluate the expectation value exactly, or at least find a way to decompose it into smaller bits. As a first step, we will find a factorization onto expectation values of polynomials labeled by connected (or indecomposable) permutations \cite{ConnectedPerm}. Then we will use the cyclic permutation invariance to reduce the number of irreducible blocks to SIF permutations \cite{SIF}.

\begin{definition}
{\rm (Decomposition into stabilized blocks).} Let $\sigma\in\mathfrak{S}_n$ and $1=i_1<i_2<\dotsb <i_p<i_{p+1}=n+1$. We say that $\{i_j\}_{j=1,\dotsc,p+1}$ decomposes $\sigma$ into (stabilized) blocks if
\be
\forall\, j \in [1,p]\qquad \sigma \left([i_j, i_{j+1}-1] \right) = [i_j, i_{j+1}-1],
\ee
i.e. $\sigma$ stabilizes the intervals $[i_j, i_{j+1}-1]$. A permutation which does not admit any block decomposition, except the trivial one ($p=1$, $i_1=1, i_2=n+1$), is called a connected permutation.
\end{definition}

\begin{definition} \label{def:ConnectedBlocks}
{\rm (Decomposition into connected blocks).} Let $\sigma\in\mathfrak{S}_n$ and $1=i_1<i_2<\dotsb <i_p<i_{p+1}=n+1$. We say that a block decomposition $\{i_j\}_{j=1,\dotsc,p+1}$ decomposes $\sigma$ into connected blocks if it contains any other block decomposition $\{i'_j\}$ as a subset, $\{i'_j\} \subset \{i_j\}$. If $\sigma$ has a decomposition $\{i_j\}_{j=1,\dotsc,p+1}$ into connected blocks, the block permutations $\sigma_j$, defined as
\be
\sigma_j(k) = \sigma(k+i_j-1),\qquad \forall\, k\in[1,i_{j+1}-i_j], j = 1,\dotsc,p,
\ee
are connected permutations.
\end{definition}

The Definition \ref{def:ConnectedBlocks} makes sense because when $\sigma$ stabilizes two intervals, it also stabilizes their intersection. Therefore the decomposition into connected blocks is obtained as
\be
\{i_j\} = \bigcup_{\{i'_k\}} \{i'_k\},
\ee
and it is unique. In practice, it can be conveniently visualized using Murasaki diagrams \cite{SIF}.

We are interested in the set $\mathcal{M}_\sigma$ of meandric systems entering the Wick expansion of $\langle P_\sigma(T, \bar{T}) \rangle$. When $\sigma\in\mathfrak{S}_n$ has a connected block decomposition $\{i_j\}_{j=1,\dotsc,p+1}$, the set of (ordered) vertices on the horizontal line has a canonical decomposition into regions
\be
\mathcal{I}_j = (i_{j\bullet},i_{j\circ}, \dotsc,(i_{j+1}-1)_\bullet, (i_{j+1}-1)_\circ).
\ee
When there is an upper arch which connects a white vertex to a black vertex $k_\bullet\in\mathcal{I}_j$, then there is a lower arch connecting the same white vertex to another black vertex $\sigma^{-1}(k)_\bullet\in\mathcal{I}_j$ in the same region (possibly $k_\bullet$ itself if it is a fixed point of $\sigma$).

There are meandric systems whose loops are each restricted to a single region and never visit two or more of them. Those meandric systems have the properties that simply cutting the horizontal line between each region, i.e. after each vertex $(i_{j+1}-1)_\circ$, $j\in[1,p-1]$, reduces them to $p$ disconnected meandric systems. There is a subset of $\mathcal{M}_{\sigma}$ consisting of systems of this type,
\be
\mathcal{M}_{\sigma_1} \times \dotsb \times \mathcal{M}_{\sigma_p} \subset \mathcal{M}_{\sigma}.
\ee
The number of such systems is the product of the number of systems in each region $\mathcal{I}_j$ where the permutation which relates the upper arch configuration to the lower configuration is $\sigma_j$. This implies the obvious bound
\be
\langle P_\sigma(T, \bar{T}) \rangle \geq \prod_j \langle P_{\sigma_j}(T,\bar{T}) \rangle,
\ee
These meandric systems which get disconnected after one cut of horizontal line are called \emph{1-reducible} in \cite{ArchStat}.

The question is then whether there are really many more than the 1-reducible meandric systems of $\mathcal{M}_{\sigma_1} \times \dotsb \times \mathcal{M}_{\sigma_p}$ contributing to $\langle P_\sigma(T, \bar{T}) \rangle$. The following Theorem shows that it barely is the case.

\begin{theorem} \label{thm:Factorization}
Let $\sigma\in\mathfrak{S}_n$ and $1=i_1<i_2<\dotsb <i_p<i_{p+1}=n+1$ such that $\{i_j\}_{j=1,\dotsc,p+1}$ decomposes $\sigma$ into connected blocks and let $\{\sigma_j\}_{j=1,\dotsc,p}$ be the corresponding connected permutations. Let $\Pl\mathfrak{S}_p$ denote the set of planar permutations on $\{1,\dotsc,p\}$.

There is a bijective map between $\left(\mathcal{M}_{\sigma_1} \times \dotsb \times \mathcal{M}_{\sigma_p}\right)\times \Pl\mathfrak{S}_p$ and $\mathcal{M}_\sigma$, which implies
\be
\langle P_\sigma(T, \bar{T}) \rangle = C_p\,\prod_{j=1}^p \langle P_{\sigma_j}(T, \bar{T}) \rangle .
\ee
\end{theorem}

The proof proceeds with a few lemmas ; all the notations are borrowed from the Theorem \ref{thm:Factorization}.

\begin{lemma} \label{lemma:Injectivity}
There is an injective map $\left(\mathcal{M}_{\sigma_1} \times \dotsb \times \mathcal{M}_{\sigma_p}\right) \times \Pl\mathfrak{S}_p \to \mathcal{M}_\sigma$. 
\end{lemma}

\begin{figure}
\subfloat[A collection of five meandric systems trivially glued together. We only draw explicitly the arches with the last white vertex of each block as a foot.]{\includegraphics[scale=.6]{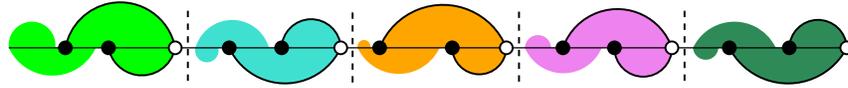}\label{fig:ConnBlocks}}\\
\subfloat[A planar permutation $\rho$ on the set of blocks of the above meandric systems. We can consider the white vertices to correspond to the last white vertex of each block.]{\includegraphics[scale=.6]{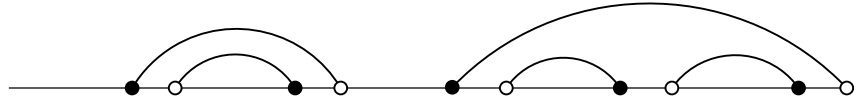}\label{fig:PlanarPerm}}\\
\subfloat[We use $\rho$ to re-arrange the arches touching the last white vertex of each block. Clearly the use of a \emph{planar} permutation prevents the new arches to cross each other. Since they only have the last white vertex of each region as white foot, they do not intersect the arches contained in each region.]{\includegraphics[scale=.6]{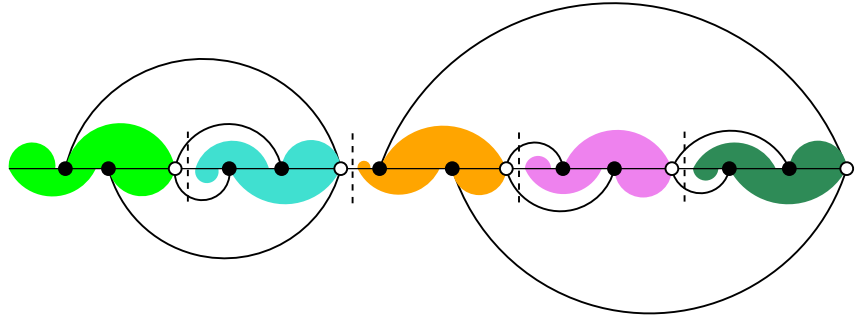}\label{fig:NewMeander}}
\caption{\label{fig:Injectivity} From a succession of disconnected meandric systems and a planar permutation on the blocks, we get a new meandric system.}
\end{figure}

{\bf Proof.} We consider $p$ meandric systems in $\mathcal{M}_{\sigma_1} \times \dotsb \times \mathcal{M}_{\sigma_p}$ and glue them together so as to obtain a meandric system of order $n$ in $\mathcal{M}_\sigma$, like in the Figure \ref{fig:ConnBlocks}. We label the vertices $(1_\bullet, 1_\circ,\dotsc,n^1_{1\bullet}, n^1_{1\circ}, 1_\bullet^2, 1_\circ^2, \dotsc,n^p_{p\bullet}, n^p_{p\circ})$ from left to right, so that $(1^j_{\bullet}, 1^j_\circ,\dotsc,n^j_{j\bullet}, n^j_{j\circ})$ are the vertices of the region $\mathcal{I}_j$.

Let $\rho\in\Pl\mathfrak{S}_p$ be a planar permutation, like in the Figure \ref{fig:PlanarPerm}, which we are going to use to create a new meandric system in $\mathcal{M}_\sigma$. All upper arches with a white vertex $k^j_{j\circ}\in\mathcal{I}_j$ different from the last vertex of the region $n^j_{j\circ}\in\mathcal{I}_j$ as a foot are left unchanged.

The $p$ remaining arches connect for each $j=1,\dotsc,p$ the white vertex $n^j_{j\circ}$ to a black vertex $l_{j\bullet}$. We cut them and rearrange them so that $l_{j\bullet}$ is now connected to $n^{\rho(j)}_{\rho(j)\circ}$, like in the Figure \ref{fig:NewMeander}. This new arch configuration is planar.
\begin{itemize}
\item Since we started from arch configurations each restricted to a region $\mathcal{I}_j$ with $n^j_{j\circ}$ as its last vertex, there is no arch going above the one between $l_{j\bullet}$ and $n^j_{j\circ}$, see the Figure \ref{fig:ConnBlocks}. Therefore when it is cut to create two new arches connected those vertices to other regions, the newly created arches do not cross any of the arches restricted to the regions $\mathcal{I}_j$.
\item Since $\rho$ is planar, it induces a planar arch configuration for the newly created arches which therefore do not cross each other.
\end{itemize}
The upper arch configuration induces a lower configuration which for the same exact reasons is planar too.

Moreover, given an arch configuration with arches possibly connecting $n^j_{j\circ}$ to other regions and all other arches connecting black to white vertices of the same region, we can reconstruct a unique element of $\left(\mathcal{M}_{\sigma_1} \times \dotsb \times \mathcal{M}_{\sigma_p}\right) \times \Pl\mathfrak{S}_p$. If $n^j_{j\circ}$ is connected to a region $\mathcal{I}_q$, $q\neq j$, there is a single black vertex $l_{j\bullet}$ also connected outside of $\mathcal{I}_j$. Then we cut those arches to connect $l_{j\bullet}$ to $n^j_{j\circ}$. This leads to a meandric system in $\mathcal{M}_{\sigma_1} \times \dotsb \times \mathcal{M}_{\sigma_p}$. The permutation $\rho\in\Pl\mathfrak{S}_p$ is found as the one which sends $n^j_{j\circ}$ to the white vertex $l_{j\bullet}$ was initially connected to.
\qed

It remains to show that the map introduced above is surjective.

\begin{lemma} \label{lemma:ConnectedPerm}
Let $\rho\in\mathfrak{S}_n$ be a connected permutation. Then,
\be
\forall k\in[1,n-1]\qquad \exists\ l>k, \quad \rho(l)\leq k.
\ee
\end{lemma}

{\bf Proof.} If there exists $k\in[2,n-1]$ such that for all $l>k$, $\rho(l)>k$, then the interval $[k+1,n]$ is stabilized and $\rho$ is not connected.
\qed

\begin{lemma} \label{lemma:LastVertexArch}
With the same hypotheses as in the Theorem \ref{thm:Factorization}, for each meandric system in $\mathcal{M}_\sigma$ and for each region $\mathcal{I}_j$, there is at most one upper arch which connects a white vertex of $\mathcal{I}_j$ to a different region, and the white vertex can only be the last vertex of $\mathcal{I}_j$, $(i_{j+1}-1)_\circ$.
\end{lemma}

{\bf Proof.} Assume that there is an upper arch which connects $k_\circ$, with $k\in[i_j, i_{j+1}-2]$, to a black vertex $\pi(k)_\bullet$ in a different region $\mathcal{I}_q$, $q\neq j$. For definiteness, we assume that $q>j$ so that $\mathcal{I}_q$ lies to the right of $\mathcal{I}_j$. Then there is also a lower arch which connects $k_\circ$ to a black vertex $\sigma^{-1}(\pi(k))_\bullet$ of $\mathcal{I}_q$,
\begin{equation*}
\begin{array}{c} \includegraphics[scale=.8]{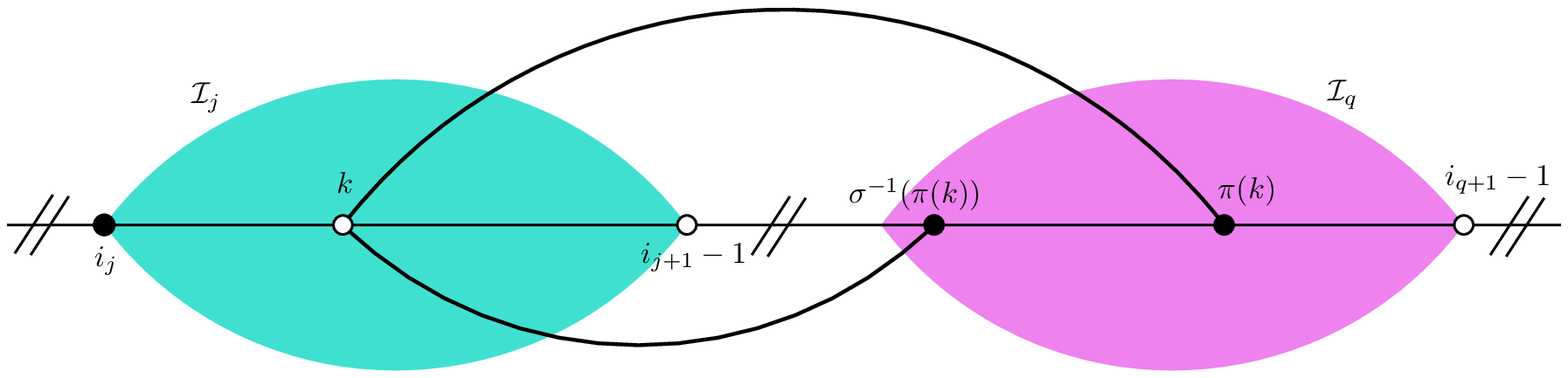} \end{array}
\end{equation*}

Applying the Lemma \ref{lemma:ConnectedPerm} to the restriction of $\sigma^{-1}$ to $[i_j,i_{j+1}-1]$, it is found that there exists a black vertex $l_\bullet$ with $k<l \leq i_{j+1}-1$, whose image $\sigma^{-1}(l)_\bullet$ is on the left of $k_\circ$, i.e. $i_j\leq \sigma^{-1}(l)\leq k$,
\begin{equation*}
\begin{array}{c} \includegraphics[scale=.8]{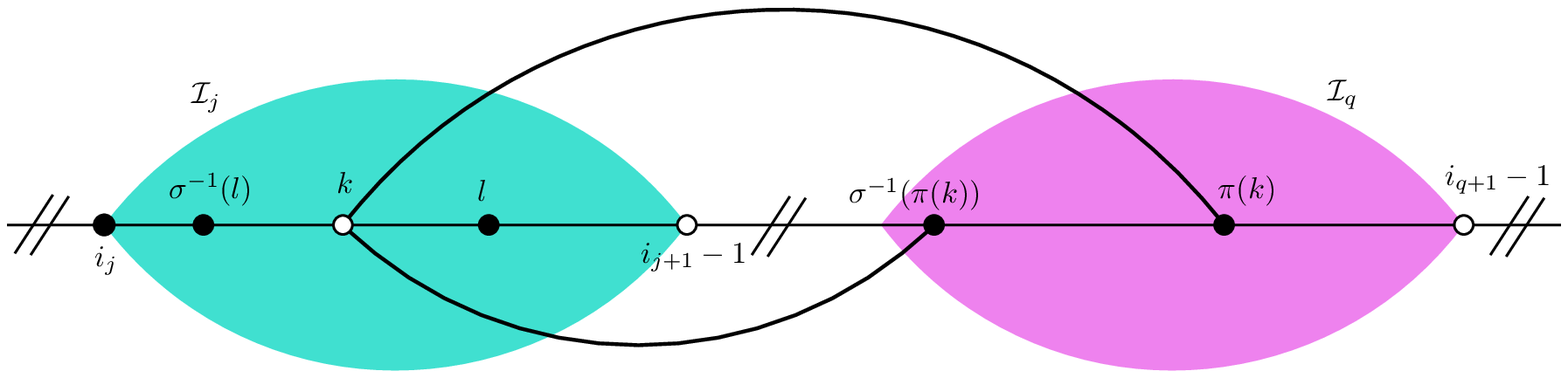} \end{array}
\end{equation*}

Now we look for the white vertices which can be connected to $l_\bullet$ via an upper arch.
\begin{itemize}
\item For planarity reason, there can be no upper arch between any white vertex on the left of $k_\circ$ and $l_\bullet$ since it would cross the arch between $k_\circ$ and $\pi(k)_\bullet$.
\item Similarly, no upper arch can connect $l_\bullet$ to any white vertex on the right of $\pi(k)_\bullet$.
\item If there is an upper arch between $l_\bullet$ and a white vertex in $[(k+1)_\circ, (i_{j+1}-1)_\circ]$, then there is a lower arch between this white vertex and $\sigma^{-1}(l)_\bullet$ which would cross the lower arch between $k_\circ$ and $\sigma^{-1}(\pi(k))$.
\end{itemize}

Therefore, the only possibility is to draw an upper arch between $l_\bullet$ and a white vertex $m_\circ\in[i_{j+1 \circ}, (\pi(k)-1)_\circ]$. To avoid a crossing in the lower plane, we further must have $\sigma^{-1}(\pi(k))\leq m < \pi(k)$. Since both $\sigma(\pi(k))_\bullet,\pi(k)_\bullet\in\mathcal{I}_q$, we find that
\begin{gather} \label{NotLastWhiteVertex}
m_\circ\in\mathcal{I}_q,\quad \text{with $m<i_{q+1}-1$},\\
\begin{array}{c} \includegraphics[scale=.8]{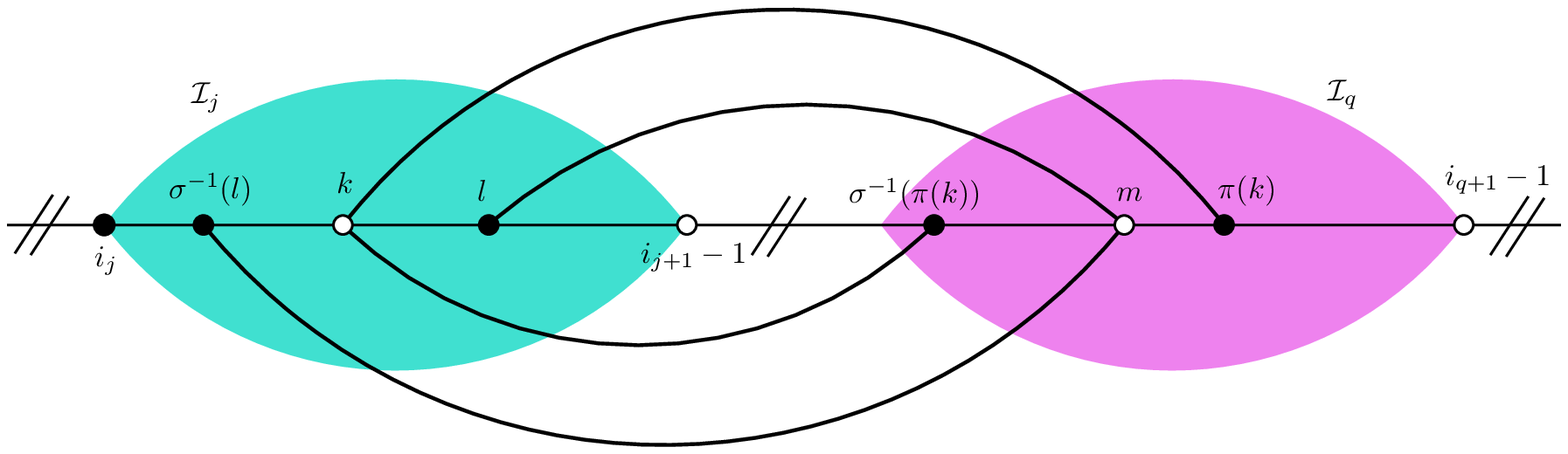} \end{array} \nonumber
\end{gather}
i.e. $m_\circ$ is \emph{not} the last white vertex of $\mathcal{I}_q$. The order of the so far relevant vertices is
\be
i_j \leq \sigma^{-1}(l)\leq k < l \leq i_{j+1}-1 < \sigma^{-1}(\pi(k)) \leq m < \pi(k)\leq i_{q+1}-1,\qquad \text{with $l=\pi(m)$}.
\ee

Thanks to \eqref{NotLastWhiteVertex} we can apply the Lemma \ref{lemma:ConnectedPerm} to the restriction of $\sigma^{-1}$ to $\mathcal{I}_q$ to get that
\begin{gather}
\exists\ r\in[i_q, m]\qquad \sigma^{-1}(r)>m,\\
\begin{array}{c} \includegraphics[scale=.8]{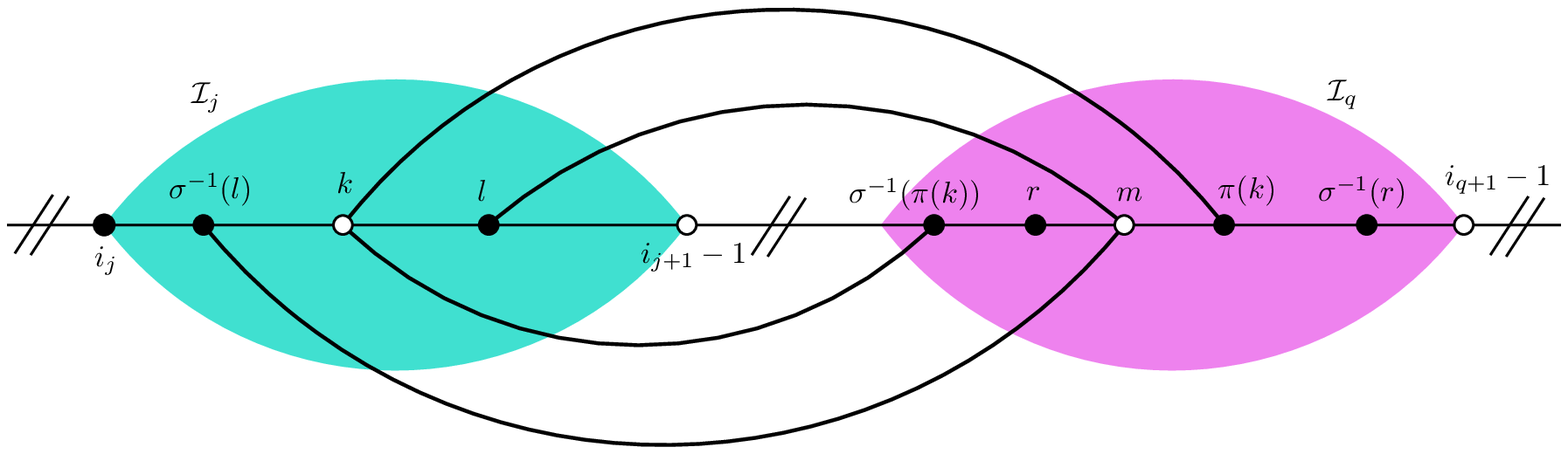} \end{array} \nonumber
\end{gather}
As it is clear from the picture, planarity requires that the white vertex connected to $r_\bullet$ by an upper arch is in $[l_\circ,(m-1)_\circ]$. This in turn gives rise to a lower arch between this white vertex and $\sigma^{-1}(r)_\bullet >m_\bullet$. This clearly breaks planarity in the lower plane,
\begin{equation*}
\begin{array}{c} \includegraphics[scale=.8]{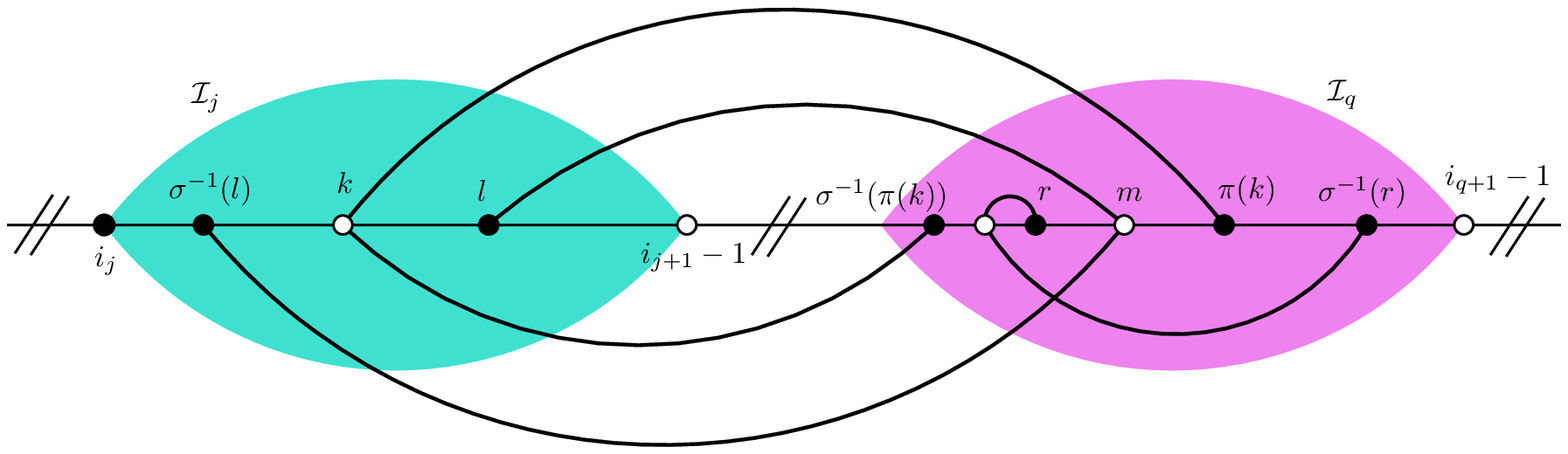} \end{array}
\end{equation*}

Consequently, there can not be any upper arch connecting $[i_{j\circ}, (i_{j+1}-2)_\circ]$ to another region.
\qed

\medskip

{\bf Proof of the Theorem \ref{thm:Factorization}.} The Lemma \ref{lemma:LastVertexArch} implies that all meandric systems in $\mathcal{M}_\sigma$ are also in the image of the map introduced in the Lemma \ref{lemma:Injectivity} and its proof. Therefore this map is also surjective, which proves the Theorem \ref{thm:Factorization}.
\qed

\medskip

Thanks to the Theorem \ref{thm:Factorization}, we are left with the problem of determining the expectation values of polynomials labeled by connected permutations. This is however not helpful in the limit of large number of vertices since the number of connected permutations behave as $n!$, \cite{SIF, ConnectedPerm}. Nevertheless, the Theorem \ref{thm:Factorization} may still apply to some connected permutations, using the fact that a cyclic re-ordering of the labels does not change the expectation value while it can turn a connected permutation in a non-connected one. Eventually one is left with \emph{stabilized-interval-free} permutations whose set of meandric systems $\{\mathcal{M}_\sigma\}_{\sigma\in\SIF_n}$ actually corresponds to the set of 2-irreducible meandric systems.

\begin{definition} \label{def:SIF}
{\rm (Stabilized-Interval-Free permutations).} We say that a permutation $\sigma\in\mathfrak{S}_n$ is stabilized-interval-free (SIF) if it does not stabilize any subinterval of $[1,n]$, i.e.
\be
\forall\ a\leq b \in [1,n] \qquad \sigma([a,b]) \neq [a,b],
\ee
except $[a,b]=[1,n]$. We denote $\SIF_n\subset \mathfrak{S}_n$ the set of SIF permutations.
\end{definition}

\begin{definition} \label{def:IrredMeanders}
{\rm (2-Reducible and irreducible meandric systems.)} We say that a meandric system is 1-reducible if a single cut on the horizontal line can produce two disconnected systems, and we say that it is 1-irreducible otherwise.
A meandric system is said to be 2-reducible if it becomes disconnected after two cuts of the horizontal line, and 2-irreducible otherwise.
\end{definition}

\begin{figure}
\includegraphics[scale=.55]{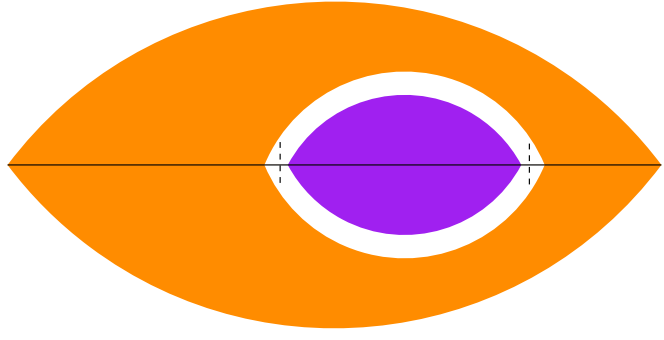}
\caption{A 2-reducible meandric systems has a sub-system totally restricted to an inner region and another sub-system which avoids this region. The two vertical dashed lines indicate the two cuts which disconnect the two meandric sub-systems.\label{fig:2Reducible}}
\end{figure}

This notion was introduced in \cite{IrredMeanders} (see also \cite{ArchStat}). A 2-reducible meandric system has the structure of the Figure \ref{fig:2Reducible}.

The Theorem \ref{thm:Factorization} has the following extension.

\begin{theorem} \label{thm:SIFFactorization}
The expectation value $\langle P_\sigma (T,\bar{T}) \rangle$ can be factorized as a product of Catalan numbers and expectation values of polynomials labeled by SIF permutations. The set $\{\mathcal{M}_\sigma\}_{\sigma\in\SIF_n}$ is the set of 2-irreducible meandric systems.
\end{theorem}

{\bf Proof.} Assume that $\sigma$ has a connected block decomposition with block permutations $\sigma_1,\dotsc,\sigma_p$. If all of them are SIF, there is nothing to prove. Assume that $\sigma_j$ for some $j$ is connected but not SIF, which means there exists a stabilized interval $[a,b]$. Then $\Sigma_j = \Delta_{-a+1}\circ \sigma_j \circ \Delta_{a-1}$ is not connected since it stabilizes $[1,b-a+1]$. Furthermore this is just a cyclic shift of the labels induced by the face of colors $(12)$, so that $P_{\sigma_j}(T,\bar{T}) = P_{\Sigma_j}(T, \bar{T})$, hence the Theorem \ref{thm:Factorization} applies to $\sigma_j$. We can do so until we are left with SIF permutations only.

The second part of the theorem states the equivalence between $\{\mathcal{M}_\sigma\}_{\sigma\in\SIF_n}$ and 2-irreducible meandric systems. Consider a 2-reducible meandric system in $\mathcal{M}_\sigma$. It contains a meandric system which can be disconnected by two cuts on the horizontal line, like in the Figure \ref{fig:2Reducible}. This system either sits in a region which starts with a black vertex $(i_\bullet,i_\circ,\dotsc,j_\bullet, j_\circ)$ or with a white vertex $((i-1)_\circ,i_\bullet,\dotsc,j_\bullet)$. In both cases, the white vertices in that region are connected by upper arches to the black vertices of $\{i_\bullet,\dotsc,j_\bullet\}$, and by lower arches to their images $\{\sigma^{-1}(i)_\bullet,\dotsc,\sigma^{-1}(j)_\bullet\}$. Clearly this set is included in $\{i_\bullet,\dotsc,j_\bullet\}$, and therefore $\sigma$ stabilizes $[i,j]$.

Reciprocally, consider a permutation $\sigma\not\in\SIF_n$ which stabilizes $[a,b]$. Thanks to a cyclic relabeling of the vertex labels, we can shift this interval to the left of the horizontal line, and work with $\tilde{\sigma} = \Delta_{-a+1} \circ \sigma \circ \Delta_{a-1}$ which stabilizes $[1,i]$ ($i=b-a+1$). $\tilde{\sigma}$ is not connected, so there exist $1<n_1<\dotsb<n_{p-1}<n_{p}=n$ which decompose $\tilde{\sigma}$ into connected blocks with permutations $(\sigma_{j})_{j=1,\dotsc,p}$. There is $k$ such that $i=n_k-1$ and for simplicity we consider $k=1$. The set $\mathcal{M}_\sigma$ can be described as $\mathcal{M}_{\sigma_1}\times \dotsb \times \mathcal{M}_{\sigma_p} \times \Pl\mathfrak{S}_p$, according to the Theorem \ref{thm:Factorization}. If $1$ is a fixed point of the planar permutation $\rho$, it means that the meandric system from $\mathcal{M}_{\sigma_1}$ is contained in $\{1_\bullet,\dotsc,n_{1\circ}\}$ and no arch connects it to the other systems. This is obviously 1-reducible, and upon a cyclic relabeling, it typically becomes 2-reducible. Now we assume $\rho(1)\neq1$ which means that the white vertex $n_{1\circ}$ is connected outside $\{1_\bullet,\dotsc,n_{1\bullet}\}$. Consequently, there is another white vertex $n_{l\circ}$ outside this region which has an upper arch and a lower arch connected in $\{1_\bullet,\dotsc,n_{1\bullet}\}$. This looks like
\be
\begin{array}{c}
\includegraphics[scale=.7]{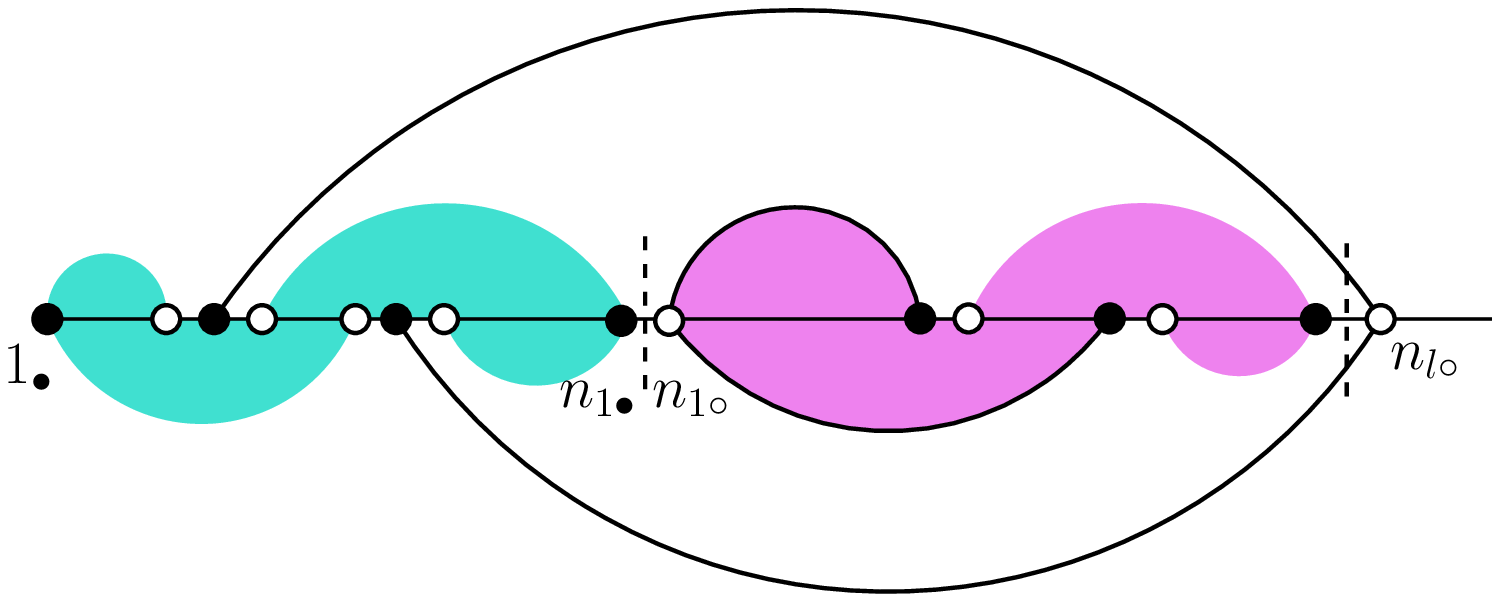}
\end{array}
\ee
The colored area represent two regions with meandric systems which can not communicate. Clearly, the two vertical dashed lines indicate places where cuts can be performed and disconnect the full system into two pieces.

\qed

\medskip

SIF permutations do not allow fixed points and therefore form a subset of the derangements. Since the number of derangements grows like $n!/e$, this is an improvement with respect to the set of connected permutations. It turns out that the number of SIF permutations also grows like $n!/e$ \cite{SIF}. A more thorough study of SIF permutations will appear in a subsequent publication.

\section{Applications and examples} \label{sec:Applications}

\subsection{Applications of the theorems}

We can use the Theorems \ref{thm:Factorization} and \ref{thm:SIFFactorization} to calculate some expectation values easily, and in particular recover analytically the number of meandric systems $M_n^{(k)}$ with $k$ reasonably close to $n$. The limitation is that we have to exhaust all the permutations on $[1,n]$ with exactly $k$ cycles before summing the corresponding expectation values.

\begin{proposition} \label{prop:NumberMeandricSys}
The numbers of meandric systems of order $n$ with $k=n, n-1, n-2$ components are
\begin{subequations}
\begin{align}
M_n^{(n)} &= C_n, \label{nCycles}\\
M_n^{(n-1)} &= n(C_{n+1}-2C_n), \label{(n-1)Cycles}\\
M_n^{(n-2)} &= n \left( C_{n+3} + \frac{3n-35}{6}\,C_{n+2} - \frac{6n-25}{3}\,C_{n+1} + 2(n-1)\,C_n\right). \label{(n-2)Cycles}
\end{align}
\end{subequations}
\end{proposition}

{\bf Proof.} \emph{Equation \eqref{nCycles}.} The only permutation with $n$ cycles is the identity whose connected block decomposition consists of $n$ blocks which are the identity on the 1-element sets $\{i\}$, $i=1,\dotsc,n$. The Theorem \ref{thm:Factorization} gives the expected well-known answer $C_n$.

\emph{Equation \eqref{(n-1)Cycles}.} The permutations with exactly $n-1$ cycles are the transpositions $\tau_{ab}$, for $1\leq a<b\leq n$. Since the expectation value is invariant under conjugation by the cyclic shift $\Delta_a$, we can consider the transposition between $1$ and $b-a+1$ instead. It has one connected block $\tau \in \mathfrak{S}_{b-a+1}$ with cycle decomposition
\be
\tau = (1\ b-a+1)(2)\dotsb (b-a),
\ee
and $n-(b-a+1)$ blocks which are the identity on $\{i\}$, $i=b-a+2, \dotsc,n$. The Theorem \ref{thm:Factorization} gives
\be
\langle P_{\tau_{ab}} \rangle = C_{n-(b-a)}\ \langle P_{\tau} \rangle.
\ee
The expectation value for $\tau$ is the same as for the transposition between the first two elements, $\tilde{\tau} = (12) (3)\dotsb (b-a+1)$, for which the Theorem \ref{thm:Factorization} yields $\langle P_{\tilde{\tau}} \rangle = 2 C_{b-a}$, as $\langle P_{(12)} \rangle =2$ for the transposition on two elements. Therefore
\be
\langle P_{\tau_{ab}} \rangle = 2\,C_{n-(b-a)}\ C_{b-a}.
\ee
To perform the sum over all transpositions, we use a reasoning that we will later reproduce in more complicated situations. The sum over $(a,b)$ can be organized as a sum over the gap $x=b-a$ and a sum over the position of $a=1,\dotsc,n$ implemented using the conjugation by $\Delta_k$ on $\tau_{1\, x+1}$, for $k=0,\dotsc,n-1$,
\be
\sum_{a<b} \langle P_{\tau_{ab}} \rangle = \alpha \sum_{x=1}^{n-1} \sum_{k=0}^{n-1} \langle P_{\Delta_{-k} \circ \tau_{1\, x+1} \circ \Delta_k} \rangle.
\ee
Here $\alpha$ is a symmetry factor which corrects for the fact that each transposition $\tau_{ij}$ appears twice in the orbit of $\tau_{1\, j-i+1}$ under the action of $(\Delta_k)_{k=0,\dotsc,n-1}$ (once with $a=i, b=j$ and once with $b=i, a=j$), hence $\alpha=1/2$. Moreover, since the action of $\Delta_k$ leaves the expectation values invariant, there are $n$ equivalent positions for $a$, which means that we can fix $a=1$ ($k=0$) and extract a factor $n$. Thus,
\be
M_n^{(n-1)} = \sum_{\tau_{ab}} \langle P_{\tau_{ab}} \rangle = \frac{n}{2} \sum_{x=1}^{n-1} \langle P_{\tau_{1\, x+1}} \rangle = \frac{n}{2} \sum_{x=1}^{n-1} 2\,C_{n-x}\,C_{x}.
\ee
Using the standard recursion $C_{j+1} = \sum_{l=0}^j C_{j-l} C_l$, we get
\be
M_n^{(n-1)} = n\, (C_{n+1}-2\,C_n).
\ee

\emph{Equation \eqref{(n-2)Cycles}.} We distinguish three types of contributions to permutations with exactly $n-2$ cycles.
\begin{enumerate}
\item \label{TwoCrossingTranspositions} $\tau_{ab}\circ\tau_{cd}$ has $n-4$ fixed points and transposes $a$ with $b$ and $c$ with $d$, \emph{with} a crossing, e.g. $1\leq a<c<b<d\leq n$.
\item \label{TwoNonCrossingTranspositions} $\tau_{ab}\circ\tau_{cd}$ has $n-4$ fixed points and transposes $a$ with $b$ and $c$ with $d$, \emph{without} crossing, e.g. $1\leq a<b<c<d\leq n$.
\item \label{3Cycle} $\sigma_{abc}$ has $n-3$ fixed points and contains a 3-cycle $(abc)$. There are two orientations for the cycle, but both leads to the same expectation value.
\end{enumerate}

\emph{Case \ref{TwoCrossingTranspositions}.} The Theorem \ref{thm:Factorization} gives
\be
\langle P_{\tau_{ab}\circ\tau_{cd}|_{a<c<b<d}} \rangle = 4\ C_{n-(d-a)}\,C_{d-b}\,C_{b-c}\,C_{c-a},
\ee
where $4$ is the expectation value for the permutation $\sigma=(13)(24)$.

The sum over $a,b,c,d$ is organized as a sum over the gaps $d-b, b-c, c-a$ and a sum over the position of $a$ implemented via the conjugation by $\Delta_k$, $k=0,\dotsc,n-1$. Since a given permutation $\tau_{ij}\circ\tau_{kl}$ will appear $4$ times in an orbit ($a=i, c=k, b=j, d=l$ and the three cyclic permutations on $a,b,c,d$, e.g. $a=k, c=j, b=l, d=i$), the symmetry factor is $\alpha=1/4$. Conjugations by $\Delta_k$ leave the expectation values invariant, meaning that there are $n$ equivalent positions for $a$. Setting $a=1$ and factorizing $n$ we get
\be
\sum_{a<c<b<d} \langle P_{\tau_{ab}\circ\tau_{cd}} \rangle = \frac{n}{4} \sum_{d=4}^{n} \sum_{b=3}^{d-1} \sum_{c=2}^{b-1} 4\ C_{n+1-d}\,C_{d-b}\,C_{b-c}\,C_{c-1}.
\ee
Using $\sum_{c=2}^{b-1} C_{b-c} C_{c-1} = C_b -2C_{b-1}$, then $\sum_{b=3}^{d-1} C_{d-b} C_b = C_{d+1} -2C_d-C_{d-1}-2C_{d-2}$ and $\sum_{b=3}^{d-1} C_{d-b} C_{b-1} = C_d-2C_{d-1} -C_{d-2}$, we arrive at
\be
\sum_{a<c<b<d} \langle P_{\tau_{ab}\circ\tau_{cd}} \rangle = n \sum_{d=4}^{n} C_{n+1-d}\,(C_{d+1}-4C_d +3C_{d-1}).
\ee
We therefore have to evaluate for $p=\pm1, 0$,
\be
\sum_{d=4}^n C_{n+1-d}\,C_{d+p} = C_{n+2+p} - 2C_{n+1+p} - \sum_{k=1}^{p+3} C_{n+1+p-k}\,C_k,
\ee
where the number of terms in the last sum is independent of $n$. This finally leads to
\be
\sum_{a<c<b<d} \langle P_{\tau_{ab}\circ\tau_{cd}} \rangle = n\,\left(C_{n+3} - 6\,C_{n+2} + 10\,C_{n+1} - 4\,C_n\right).
\ee

\emph{Case \ref{TwoNonCrossingTranspositions}.} There are two typical patterns, one where the two transpositions are separated, e.g. $a<b<c<d$, and the other where they are nested, e.g. $d<a<b<c$. In the nested case we have
\be
\langle P_{\tau_{ab} \circ \tau_{cd}|_{d<a<b<c}} \rangle = 2\times 2\times C_{n-(c-d)}\,C_{c-d-(b-a)}\,C_{b-a}.
\ee
As before, we first keep the distances between the elements which are transposed fixed and sum over the position of $a$ using the orbit generated by $\Delta_k$. Notice that along an orbit one pattern can be turned into the other. This implies in particular that both patterns have the same expectation values. Therefore the symmetry factor is $\alpha=2/4$, where $2$ comes from the two patterns and $1/4$ from the number of times a permutation appears. We fix $d=1$, extract a factor $n$, and
\be
\sum_{a<b<c<d} \langle P_{\tau_{ab}\circ\tau_{cd}} \rangle + \sum_{d<a<b<c} \langle P_{\tau_{ab}\circ\tau_{cd}} \rangle = 2n \sum_{c=4}^n \sum_{b=3}^{c-1} \sum_{a=2}^{b-1} C_{n+1-c}\, C_{c-(b-a)-1}\, C_{b-a}.
\ee
The absolute positions of $a$ and $b$ are irrelevant and only the gap $x=b-a$ matters. There are $c-x-2$ possible positions for $a$, hence (after the change $c \leftarrow c-1$)
\be
\sum_{a<b<c<d} \langle P_{\tau_{ab}\circ\tau_{cd}} \rangle + \sum_{d<a<b<c} \langle P_{\tau_{ab}\circ\tau_{cd}} \rangle = 2n \sum_{c=3}^{n-1} C_{n-c} \sum_{x=1}^{c-2} (c-1-x)\,C_{c-x}\,C_x.
\ee
In the final steps, the following formula for $i<j<k$, $j<k-i$ is used several times,
\be
\begin{aligned}
\sum_{x=i}^{k-j} x\,C_{k-x}\,C_x &= \frac{1}{2} \sum_{x=i}^{k-j} x\,C_{k-x}\,C_x+ \frac{1}{2} \sum_{y=j}^{k-i} (k-y)\,C_{k-y}\,C_y,\\
&= \frac{k}{2} \sum_{x=j}^{k-j} C_{k-x}\,C_x +\frac{1}{2} \sum_{x=i}^{j-1} x\,C_{k-x}\,C_x + \frac{1}{2} \sum_{x=k-j+1}^{k-i} (k-x)\,C_{k-x}\,C_x,\\
&= \frac{k}{2}\,C_{k+1} -k \sum_{x=0}^{j-1} C_{k-x}\,C_x + \sum_{x=i}^{j-1}x\,C_{k-x}\,C_x.
\end{aligned}
\ee
On the first line, we split the sum into two halves and relabel one of them $y=k-x$, On the second line we factorized the common terms, and noticed in the third line that the sum which has a $k$-dependent number of terms can be done. The number of terms in the remaining sums is independent of $k$. A similar formula holds for $j<i$. This allows to perform the sum over $x$ and then the sum over $c$, to get
\be
\sum_{a<b<c<d} \langle P_{\tau_{ab}\circ\tau_{cd}} \rangle + \sum_{d<a<b<c} \langle P_{\tau_{ab}\circ\tau_{cd}} \rangle = \frac{n}{2} \Bigl[ (n-5)\,C_{n+2} - 2(2n-9)\,C_{n+1} +4(n-3)\,C_n\Bigr].
\ee

\emph{Case \ref{3Cycle}.} The Theorem \ref{thm:Factorization} provides the expectation values,
\be
\langle P_{\sigma_{abc}|_{a<b<c}} \rangle = \langle P_{\sigma_{acb}|_{a<b<c}} \rangle = 4\,C_{n-(c-a)}\,C_{c-b}\,C_{b-a},
\ee
with $\langle P_{(123)} \rangle = \langle P_{(132)} \rangle =4$. To sum over the permutations $\sigma_{abc}$, we again sum over the positions of $a=1,\dotsc,n$ using the action of $(\Delta_k)_{k=0,\dotsc, n-1}$ and over the gaps $b-a, c-b$. The symmetry factor is $\alpha=1/3$. Therefore,
\be
\sum_{a<b<c} \langle P_{\sigma_{abc}} \rangle + \langle P_{\sigma_{acb}} \rangle = \frac{2\times4}{3}\,n \sum_{c=3}^{n} \sum_{b=2}^{c-1} C_{n+1-c}\, C_{c-b}\, C_{b-1}.
\ee
It is then straightforward to get
\be
\sum_{a<b<c} \langle P_{\sigma_{abc}} \rangle + \langle P_{\sigma_{acb}} \rangle = \frac{8}{3}\,n\Bigl(C_{n+2}-4\,C_{n+1}+3\, C_{n}\Bigr).
\ee

Finally,
\be
M_n^{(n-2)} = \sum_{a<c<b<d} \langle P_{\tau_{ab}\circ\tau_{cd}} \rangle + \sum_{a<b<c<d} \langle P_{\tau_{ab}\circ\tau_{cd}} \rangle + \sum_{d<a<b<c} \langle P_{\tau_{ab}\circ\tau_{cd}} \rangle + \sum_{a<b<c} \langle P_{\sigma_{abc}} \rangle + \langle P_{\sigma_{acb}} \rangle,
\ee
leads to the conclusion.
\qed

\subsection{Some expectation values for SIF permutations}

In all cases the Theorems \ref{thm:Factorization}, \ref{thm:SIFFactorization} are applied, the Gaussian expectation values of polynomials labeled by SIF permutations are eventually needed. In the previous applications, they were permutations on very few elements (for $\sigma = (12), (123), (13)(24)$). Here we study a family of SIF permutations on an arbitrary number of elements.

\begin{proposition} \label{prop:SIFVEV}
Let $\Delta_k$ be the cyclic permutation $i\mapsto i+k \mod n$ on $[1,n]$. Then
\begin{subequations}
\begin{align}
&\text{For $k=0\mod n$} & & \langle P_{\Delta_{0}} \rangle = C_n,\\
&\text{For $k=\pm1\mod n$} & & \langle P_{\Delta_{\pm1}} \rangle = \Mot(n),\label{Motzkin} \\
&\text{Other cases} & & \langle P_{\Delta_{k}} \rangle = n, \label{Delta_k>2}
\end{align}
\end{subequations}
where $C_n$ and $\Mot(n)$ are the Catalan and Motzkin numbers of order $n$.
\end{proposition}

{\bf Proof.} \emph{Equation \eqref{Motzkin}.} $\Delta_{\pm1}$ has a single cycle therefore the expectation value $\langle P_{\Delta_{\pm1}} \rangle$ counts a number of meanders (a single component). For definiteness, we will only consider the case $\Delta_{-1}$, so that the relevant meanders are such that there is an upper arch connecting $i_\circ$ to $j_\bullet$ if and only if there is a lower arch connecting $i_\circ$ to $\Delta_{-1}^{-1}(j)_\bullet = (j+1)_\bullet$.

We aim at a recursion on the degree $n$ of the polynomial and for the time of the proof we switch to the better adapted notation $\langle P_{\Delta_{-1}} \rangle = m_n$ for $\Delta_{-1}\in\mathfrak{S}_n$.

Let $k\in[2,n]$ and denote $m_{n,k}$ the number of contributing meanders with an upper arch between $1_\circ$ and $k_\bullet$. They also have a lower arch connecting $1_\circ$ to $(k+1)_\bullet$,
\begin{equation*}
\begin{array}{c} \includegraphics[scale=.65]{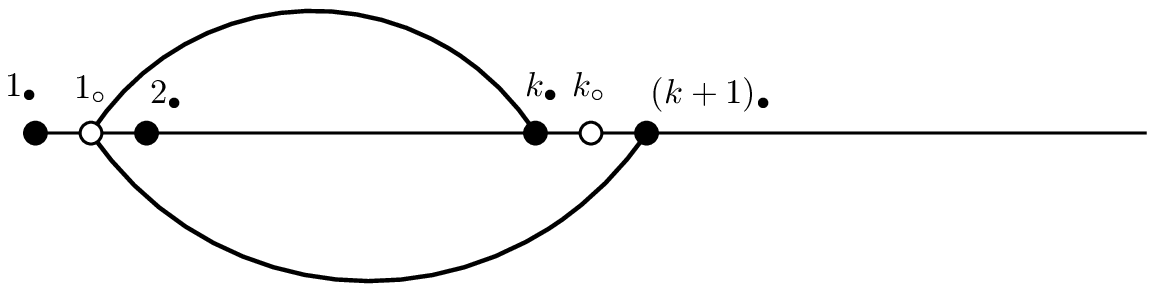} \end{array}
\end{equation*}
Then we show that the upper arch with $1_\bullet$ as a foot can only connect to $k_\circ$. Indeed, first notice that due to planarity in the upper half plane, no vertex $i_\circ$ for $i\in[2,k-1]$ can be connected to $1_\bullet$. Second, if $i_\circ$, for $i\in[k+1,n]$, is connected to $1_\bullet$, then there is a lower arch which connects $i_\circ$ to $2_\bullet$ and it clearly crosses the lower arch between $1_\circ$ and $(k+1)_\bullet$,
\begin{equation*}
\begin{array}{c} \includegraphics[scale=.65]{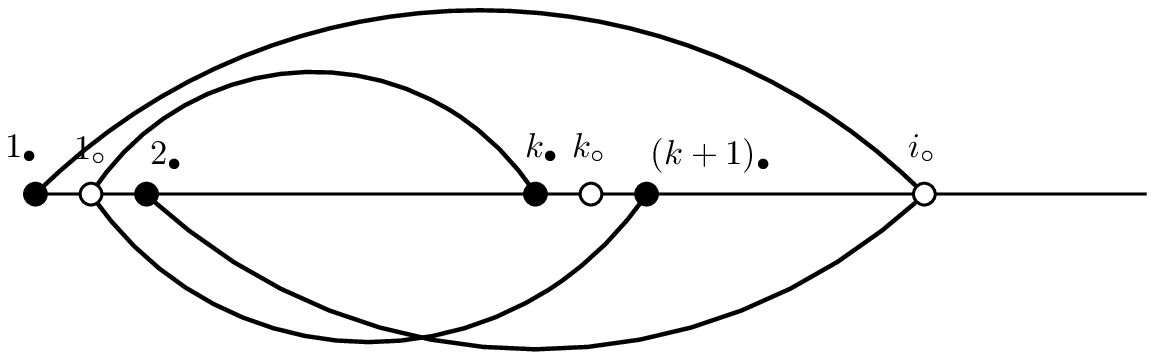} \end{array}
\end{equation*}
Therefore there is an upper arch between $1_\bullet$ and $k_\circ$ and a lower arch between $2_\bullet$ and $k_\circ$,
\begin{equation*}
\begin{array}{c} \includegraphics[scale=.65]{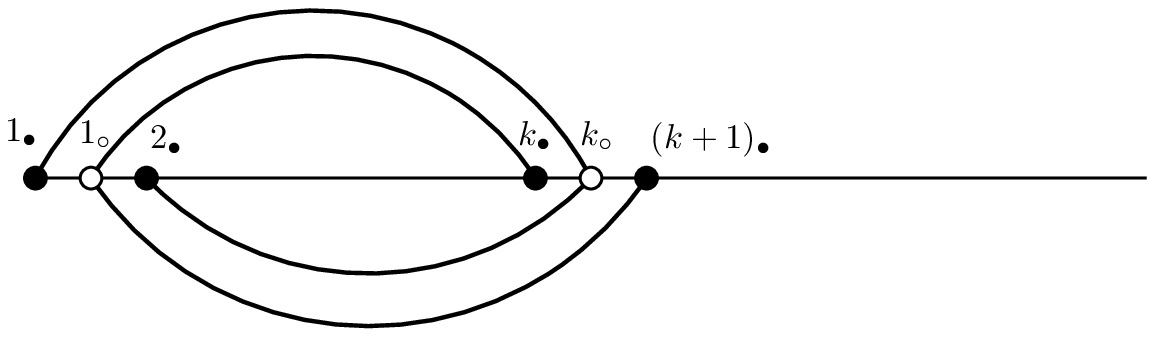} \end{array}
\end{equation*}

It is then easy to see that the number of arch systems allowed in the region $\{2_\bullet,2_\circ,\dotsc,(k-1)_\bullet, (k-1)_\bullet\}$ is precisely $m_{k-2}$. Indeed, any meanders in $\mathcal{M}_{\Delta_{-1}}$ of order $k-2$ can be inserted in the region $\{2_\bullet,2_\circ,\dotsc,(k-1)_\bullet, (k-1)_\bullet\}$ by changing the lower arch connected to $2_\bullet$ to an arch connected to $k_\bullet$ instead, and this works the other way around. Similarly, the number of arch systems allowed in the region $\{(k+1)_\bullet,(k+1)_\circ,\dotsc,n_\bullet, n_\circ\}$ is $m_{n-k}$. Consequently, for $k\in[2,n]$,
\be
m_{n,k} = m_{k-2}\,m_{n-k}.
\ee

For $k=1$, it is even simpler to find
\be
m_{n,1} = m_{n-1}.
\ee
To get to $m_n$, it only remains to sum over the position of $k$,
\be
m_n = \sum_{k=1}^n m_{n,k} = m_{n-1} + \sum_{k=2}^n m_{k-2}\,m_{n-k} = m_{n-1} + \sum_{p=0}^{n-2} m_p\,m_{n-2-p}.
\ee
Together with the initial conditions $m_0=m_1=1$, this recursion defines the Motzkin numbers and $m_n = \Mot(n)$.

\emph{Equation \eqref{Delta_k>2}.} Let $p\in[2,n-2]$. Our strategy is to prove that choosing $\pi(1)_\bullet\in [1,n]$ completely determines a meandric system. The following lemma will be useful.

\begin{lemma} \label{lemma:Delta_k>2}
Let $p\in[2,n-2]$. If a meandric system in $\mathcal{M}_{\Delta_{-p}}$ has an upper arch between $1_\circ$ and $\pi(1)_\bullet = k_\bullet$ with $k\geq4$ and $k+p\leq n+3$ , then there is also an upper arch between $2_\circ$ and $(k-1)_\bullet$.
\end{lemma}

\emph{Proof of the Lemma.} The meandric systems in $\mathcal{M}_{\Delta_{-p}}$ are such that for every upper arch between $i_\circ$ and $j_\bullet$, there is a lower arch between $i_\circ$ and $(j+p)_\bullet\mod n$, and reciprocally.

Let $k\geq4$ with $k+p\leq n+2$ for the time being, and consider an upper arch between $1_\circ$ and $k_\bullet=\pi(1)_\bullet$, together with the lower arch between $1_\circ$ and $\Delta_p(\pi(1)) = (k+p)_\bullet$,
\begin{equation*}
\begin{array}{c} \includegraphics[scale=.7]{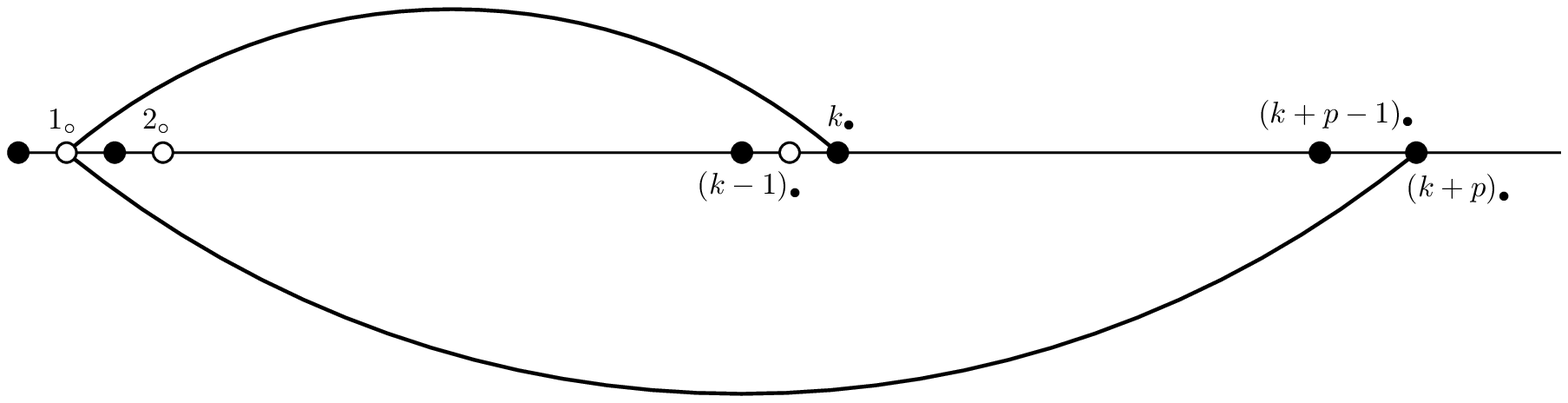} \end{array}
\end{equation*}
The drawing is made for $k+p\leq n$, but everything works the same for $k+p>n$ with $k+p-n\in\{1,2\}$.

We are going to prove that there must be an upper arch between $2_\circ$ and $(k-1)_\bullet$. Assume that in the upper half plane $2_\circ$ is connected to $j_\bullet\neq (k-1)_\bullet$, and $l_\circ\neq 2_\circ$ to $(k-1)_\bullet$, with $2\leq j\leq l\leq k-1$, $l\geq3$, $j\leq k-2$. In the lower half plane, they induce an arch between $2_\circ$ and $(j+p)_\bullet\leq (k+p-2)_\bullet$, and an arch between $l_\circ$ and $(k+p-1)_\bullet$,
\begin{equation*}
\begin{array}{c} \includegraphics[scale=.7]{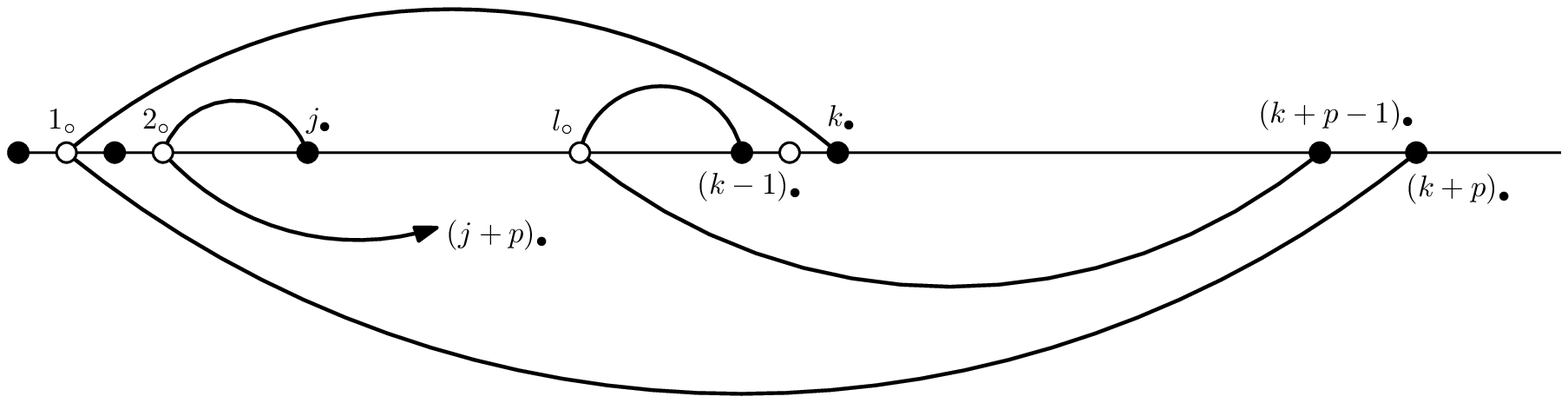} \end{array}
\end{equation*}
Given that
\be
2<l<k+p-1,\qquad\text{and}\qquad 2<j+p<k+p-1,
\ee
we find that
\be
\begin{cases}
\text{either } &2<l<j+p<k+p-1,\quad \text{then the lower arches cross each other,}\\
\text{or } &2<j+p\leq l<k+p-1,\quad \text{then they do not cross each other.}
\end{cases}
\ee
Now we focus on the second case, $j+p\leq l$,
\begin{equation*}
\begin{array}{c} \includegraphics[scale=.7]{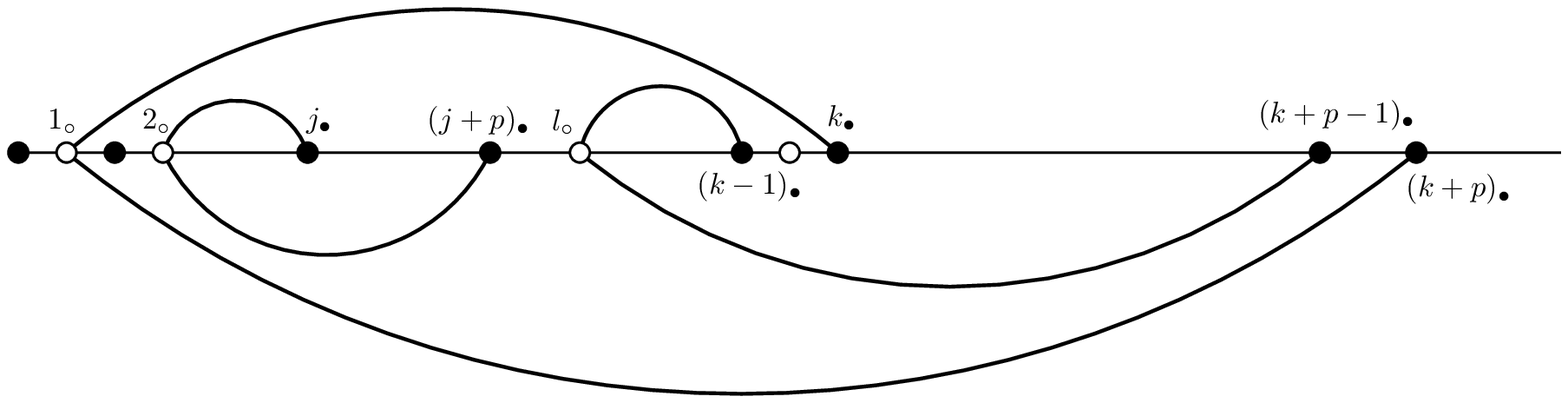} \end{array}
\end{equation*}
On the drawing, we have considered $k+p-1\leq n$, but in the case $k+p=n+2$, we have to use $\Delta_p(k-1)_\bullet = (n+1)_\bullet \mod n = 1_\bullet$. It does not change the arguments below.

Notice that due to $p\geq2$ as well as $j\geq 2$, $j+p\geq 4$. Therefore there exists a vertex $s_\circ\in[3,j+p-1]$ which is surrounded in the lower half plane by the arch between $2_\circ$ and $(j+p)_\bullet$ and must be connected to $3_\bullet$ via a lower arch,
\begin{equation*}
\begin{array}{c} \includegraphics[scale=.7]{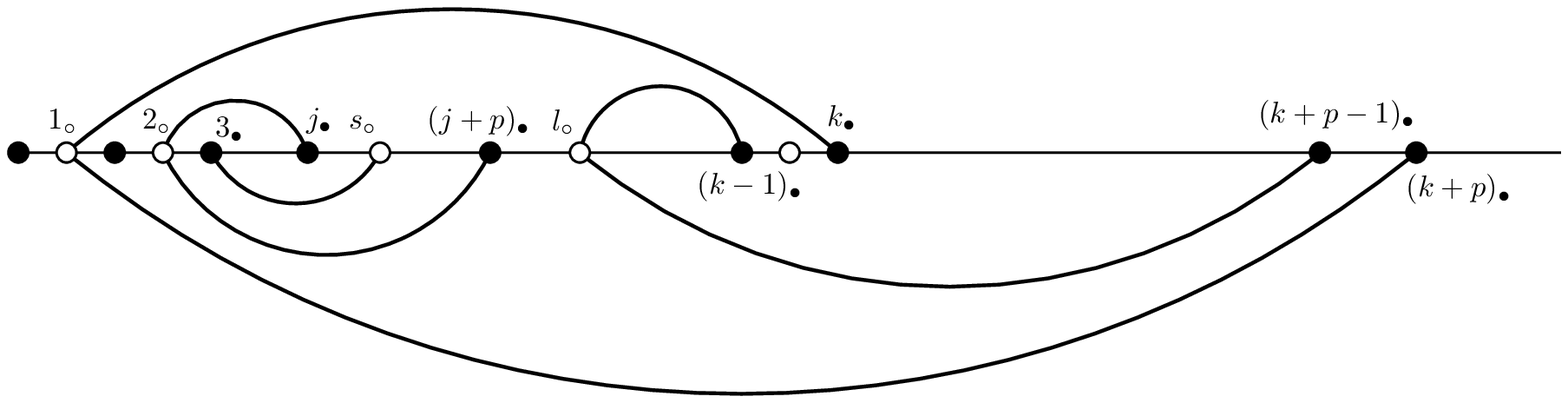} \end{array}
\end{equation*}
In the upper half plane, $s_\circ$ is thus connected to $\pi(s)_\bullet=\Delta_{-p}(3)_\bullet = (3-p)_\bullet\mod n$. If $p=2$, this is $1_\bullet$ but this cannot be planar in the upper half-plane. So we are left with the case $p\geq3$ and an upper arch between $s_\circ$ and $(n-p+3)_\bullet$. However,
\be \label{BreakPlanar}
s<k,\qquad \text{and} \qquad \pi(s) = n-p+3\geq k+1>k,
\ee
the second inequality being due to our starting assumption $k+p\leq n+2$. As a result, the upper arch between $1_\circ$ and $\pi(1)_\bullet = k_\bullet$ intersects the upper arch $s_\circ$ and $\pi(s)_\bullet=(n-p+3)_\bullet$,
\begin{equation*}
\begin{array}{c} \includegraphics[scale=.7]{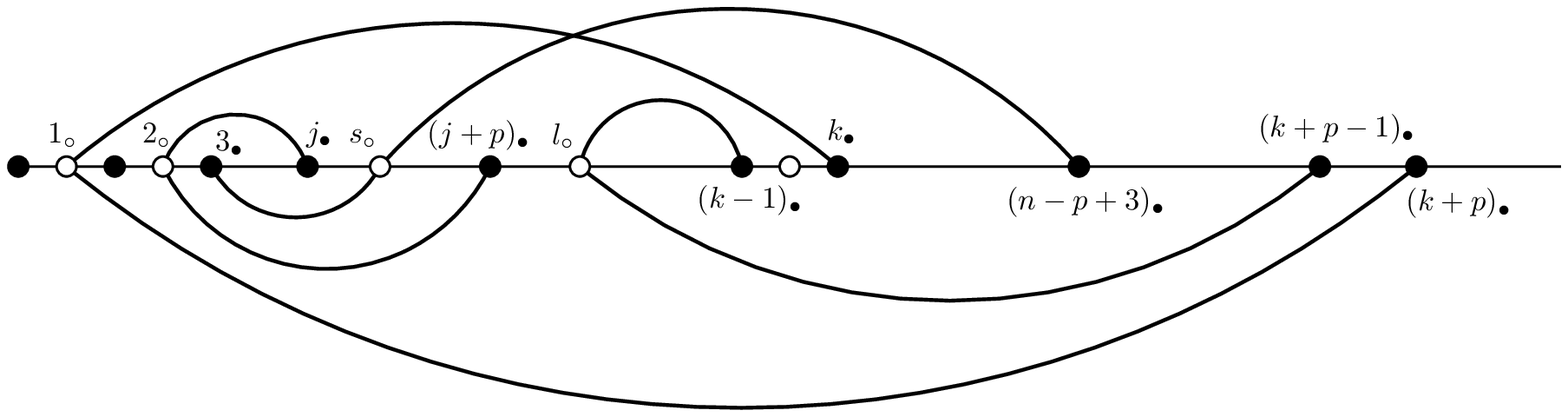} \end{array}
\end{equation*}
Therefore we must have $j=k-1$ and $l=2$.

In the case $k+p=n+3$, the Equation \eqref{BreakPlanar} gives $\pi(s)=n-p+3=k$, but $k$ is already $\pi(s)$ so this is impossible. We can also look at this case directly. If $k+p=n+3$, then $\Delta_p(k) = (k+p)\mod n = 3$, meaning that there is lower arch between $1_\circ$ and $3_\bullet$. This enforces a lower arch between $2_\circ$ and $2_\bullet$ which can not have other partners. In the upper half-plane that new arch induces an arch between $2_\circ$ and $\Delta_{-p}(2)_\bullet = (2-p+n)_\bullet = (k-1)_\bullet$ which is the expected result.
\qed

\emph{Back to the Equation \eqref{Delta_k>2}.}
\begin{itemize}
\item The Lemma \ref{lemma:Delta_k>2} directly applies to the cases $k\geq 4$ with $k+p\leq n+3$.
\item When $k+p\geq n+4$, we can simply flip the system with respect to the horizontal line, exchanging the upper and lower half-planes. This gives an upper arch between $1_\circ$ and $k'_\bullet =(k+p-n)_\bullet$ and a lower arch between $1_\circ$ and $\Delta_{n-p}(k')_\bullet= (k'+n-p)_\bullet$. Since $k'\geq4$ and $k'+n-p = k\leq n$, we can apply the Lemma \ref{lemma:Delta_k>2} to the flipped meandric system with $k'$ and $\Delta_{p-n}$.
\item When $k=2$ and $k=3$, by flipping the system again to exchange the upper and lower half-planes, we can work with the permutation $\Delta_{p-n}$ instead of $\Delta_{-p}$. Setting $k'=k+p\geq 4$, the Lemma \ref{lemma:Delta_k>2} applies since $k'+(n-p)= n+k\leq n+3$.
\item The last case to analyze is $k=1$. If $p+1\geq 4$, we can proceed just like for $k=2,3$. If $p+1=3$, there must be a lower arch between $2_\circ$ and $2_\bullet$ which implies in the upper half-plane an arch between $2_\circ$ and $n_\bullet$. Then moving the pair of vertices $(1_\bullet, 1_\circ)$ to the far right of the horizontal line, we are again in position to apply the Lemma \ref{lemma:Delta_k>2}.
\end{itemize}

The result of this analysis is that once $k_\bullet = \pi(1)_\bullet$ is chosen, there must be another arch in the upper half-plane (up to a flip of the system) below the initial one. This gives rise to a recursive process which fills a region, say $\{1_\circ,2_\bullet,\dotsc,k_\bullet\}$ (up to a redefinition of $k$), with arches on top of one another,
\begin{equation*}
\begin{array}{c} \includegraphics[scale=.7]{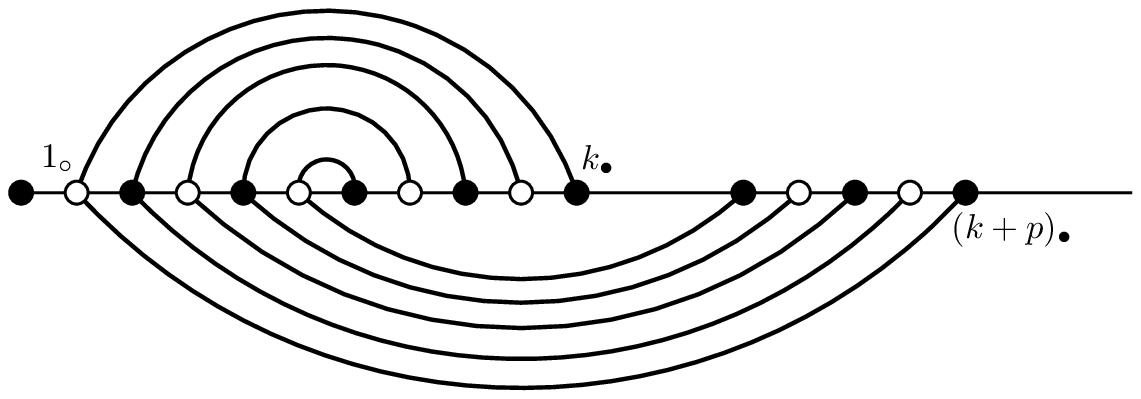} \end{array}
\end{equation*}

If $k$ is even, then we cyclically shift the vertices so that the pair $((k/2+1)_\bullet, (k/2+1)_\circ)$ becomes the leftmost pair on the horizontal line. Then the arch between $(k/2+1)_\circ$ and $(k/2)_\bullet$ becomes an arch between the first white vertex and the last black vertex of the horizontal line,
\begin{equation*}
\begin{array}{c} \includegraphics[scale=.7]{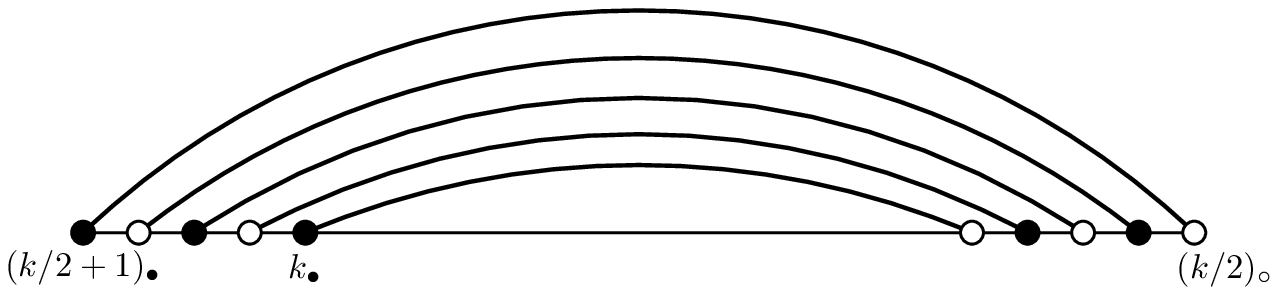} \end{array}
\end{equation*}
which means that the above recursive process applies again, until all arches are determined.

If $k$ is odd, we cyclically shift the vertices so that the pair $(((k+3)/2)_\bullet, ((k+3)/2)_\circ)$ becomes the leftmost pair on the horizontal line. The same result is eventually obtained.

Since $k$ can take $n$ values, the expectation value is simply $n$.
\qed

\bigskip

\emph{Motzkin paths.} It is well-known that planar arch configurations are one-to-one with Dyck paths. A \emph{Dyck path} of order $2n$ is a $2n$-step path in the upper half-plane which starts at $(0,0)$, ends at $(2n,0)$ and for which only two types of steps are allowed, the north-east step $(+1,+1)$ and the south-east step $(+1,-1)$. Given an arch configuration over $2n$ vertices on a horizontal line, oriented west to east, we list between each pair of consecutive vertices the number of arches which pass. This produces a list of $(2n-1)$ positive integers $(h_1,\dotsc,h_{2n-1})$ which are interpreted as the heights of a Dyck path after the step $1,\dotsc,2n-1$.

\begin{figure}
\subfloat[]{\includegraphics[scale=.65]{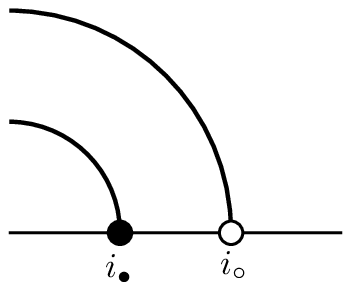}\label{fig:TwoDown}}\qquad
\subfloat[]{\includegraphics[scale=.65]{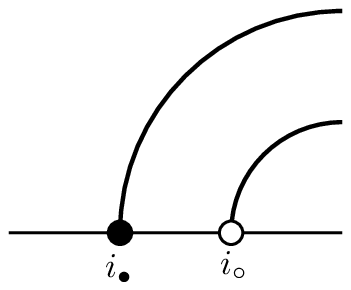}\label{fig:TwoUp}}\qquad
\subfloat[]{\includegraphics[scale=.65]{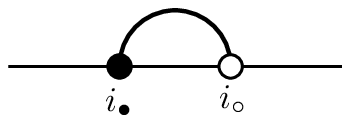}\label{fig:OneUpOneDown}}\qquad
\subfloat[]{\includegraphics[scale=.65]{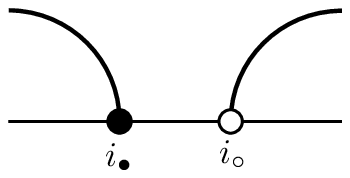}\label{fig:OneDownOneUp}}
\caption{The four possible patterns at a pair of vertices in the upper half-plane of a meandric system.\label{fig:FourPatternsDyck}}
\end{figure}
Motzkin numbers are known to count \emph{Motzkin paths}. A Motzkin path of length $n$ is a path of $n$ steps in the upper half-plane which starts at $(0,0)$ and ends at $(n,0)$ with three types of steps, north-east $(+1, +1)$, south-east $(+1, -1)$ or east $(+1, 0)$, i.e. the horizontal step.

Therefore it is interesting to find a bijection between $\mathcal{M}_{\Delta_{\pm 1}}$ and the set of Motzkin paths. Consider a meander in $\mathcal{M}_{\Delta_{-1}}$. Looking at a pair $(i_\bullet, i_\circ)$ of vertices, there are four patterns which can arise the upper half-plane, displayed in the Figure \ref{fig:FourPatternsDyck}. In terms of Dyck paths, they represent the four possible combinations of two successive steps: Figure \ref{fig:TwoDown} is two steps down, \ref{fig:TwoUp} two steps up, \ref{fig:OneUpOneDown} one up and one down, and finally \ref{fig:OneDownOneUp} one down and one up. However, meanders in $\mathcal{M}_{\Delta_{-1}}$ can not contain the pattern \ref{fig:OneDownOneUp}. Indeed, up to a cyclic permutation of the vertex labels, we can assume that we are looking at the leftmost pair of vertices $(1_\bullet, 1_\circ)$. Then, they are either connected together like in the Figure \ref{fig:OneUpOneDown}, or the white vertex $1_\circ$ is connected to some $j_\bullet$, $j>1$. In this case, we know from the proof of Equation \eqref{Motzkin} in the Proposition \ref{prop:SIFVEV} that $1_\bullet$ is connected by an upper arch to $j_\circ$. Therefore when the labels are cyclically shifted, two arches always connect the pair $(i_\bullet,i_\circ)$ to the two vertices of a pair $((j+i-1)_\bullet, (j+i-1)_\circ)$ ($\mod n$). Therefore these two arches always point in the same direction, like in the Figures \ref{fig:TwoDown}, \ref{fig:TwoUp}.

As a consequence, only three patterns in the upper-half plane are allowed. To find Motzkin paths, it is sufficient to just associate with the Figure \ref{fig:TwoDown} the south-east step, with \ref{fig:TwoUp} the north-east step and with \ref{fig:OneUpOneDown} the horizontal step. The other way around it is straightforward to show that a Motzkin path gives rise to a single meander in $\mathcal{M}_{\Delta{-1}}$.

\section*{Conclusion}

We have shown in this paper that it is possible to perform calculations of expectation values beyond the so-called melonic sector in the Gaussian random tensor model. We have considered the family of polynomials in the tensor entries whose graphical representation possesses a single face with colors $(12)$ and a single face with colors $(34)$. This generalizes the single-trace invariant of random matrix models to two faces superimposed on the same set of vertices instead of a single face. The expansion of their expectation value onto Feynman graphs is equivalent to a problem of enumeration of meandric systems whose lower and upper arch configurations are related by a permutation on the arch feet. The Theorems \ref{thm:Factorization} and \ref{thm:SIFFactorization} reduce the difficulty to the evaluation on SIF permutations \cite{SIF}, which enumerate the irreducible meandric systems of \cite{IrredMeanders} (see also \cite{ArchStat}). In the Proposition \ref{prop:SIFVEV} we have further evaluated the expectation values of polynomials labeled by some SIF permutations.

All the proofs of the paper use the meandric representation of the Feynman expansion which turns out very convenient. However, we want to stress that we could have used the set of \emph{Schwinger-Dyson Equations} (SDE) instead. This is a set of equations that is derived from the integral expression \eqref{DefVEV} (see \cite{BubbleAlgebra}) and generalizes the Tutte equation for the resummation of planar maps to Feynman amplitudes in quantum field theory. Those equations have already been used to solve tensor models at large $N$ in \cite{SDE}, and they also work in the present case. However, we have decided not to include the proofs using the SDE for two reasons: introducing them in the case of tensor models is space consuming, and the proofs would be quite redundant. Indeed, the SDE form an algebraic system on the expectation values of polynomials. Since it does not rely on the Feynman expansion, it seems at first that solving them does not involve any planarity requirement like the meandric representation. However, at large $N$ we can show that only the polynomials which maximize the number of connected components on the colors $(12)$ and $(34)$ contribute (the "freeness" property) and this is actually equivalent to planarity of the Feynman graphs. Therefore proving our results using the SDE would consist in repeating the same arguments using "maximal number of connected components of subgraphs" instead of "planarity".

Nevertheless, the question of going further than our results which rely on direct combinatorial analysis remains open and could benefit of the use of the SDE. A similar question is whether the Gaussian tensor model and its SDE can be useful to meander theory. It has been shown that random matrix models are useful to meanders \cite{Meanders:ExactAsymptotics}. As for random tensor models, we have shown that some simple results of meander theory can be recovered, like the Proposition \ref{prop:NumberMeandricSys}. But it is far from clear that more advanced results can be reproduced, like those of \cite{Meanders:ExactAsymptotics,MeanderDeterminants}. An important difference in our work is that we have enlarged the set of configurations from planar arch configurations to permutations. While this may be useful, it also means that most expectation values actually vanish since even after reduction by the Theorem \ref{thm:SIFFactorization}, we are left with SIF permutations whose number grows like $n!/e$ while the total number of meandric systems grows exponentially, $C_n^2\sim K n^{-3} (16)^n$. This is already a new piece of information for random tensor models. But it may be a drawback to progress in meander theory since it seems really difficult to find necessary and/or sufficient conditions on the permutation $\sigma$ for an expectation value $\langle P_\sigma\rangle$ to vanish\footnote{For instance, the polynomial labeled by
\be
\sigma = (1 6 3 9 7 4 8 2 5)\ \in\,\mathfrak{S}_9
\ee
in cycle notation, has a vanishing large $N$ Gaussian expectation value. It satisfies the property
\be
\forall i, i<\sigma(i), \exists j\in[i,\sigma(i)]\quad \sigma(j)\not\in [i,\sigma(i)].
\ee
and similarly for all $i>\sigma(i)$. But it is not a sufficient condition since the permutation $(1 5 2 8 6 4 7 3)\in \mathfrak{S}_8$ also satisfies this property but has a non zero expectation.}.


\section*{Acknowledgements}

Both authors are thankful to Perimeter Institute for Theoretical Physics which made this collaboration possible.


\end{document}